\documentclass{jfm}

\newcommand{\bsub}{\begin{subequations}}
	\newcommand{\esub}{\end{subequations}}

\newcommand{\al}{\alpha}

\newcommand{\ep}{\epsilon}

\newcommand{\pxi}{\mbox{\boldmath $\xi$}}

\newcommand{\pF}{\textbf{\emph{F}}}

\newcommand{\pX}{\textbf{\emph{X}}}

\newcommand{\pV}{\textbf{\emph{V}}}

\newcommand{\pps}{\textbf{\emph{s}}}

\newcommand{\pu}{\textbf{\emph{u}}}
\newcommand{\pv}{\textbf{\emph{v}}}

\newcommand{\px}{\textbf{\emph{x}}}

\newcommand{\pat}{\partial}

\newcommand{\x}{\times}

\newcommand{\beq}{\begin{equation}}
\newcommand{\eeq}{\end{equation}}
\newcommand{\bsubeq}{\begin{subequations}}
	\newcommand{\esubeq}{\end{subequations}}
\newcommand{\beqn}{\begin{eqnarray}}
\newcommand{\eeqn}{\end{eqnarray}}
\newcommand{\fr}{\frac}
\newcommand{\lb}{\label}
\newcommand{\er}{\eqref}

\usepackage{hyperref}
\hypersetup{colorlinks,allcolors=blue}
\usepackage{amsmath}
\usepackage{multirow}
\usepackage{amsfonts,amssymb}
\usepackage{graphicx}
\usepackage{subfigure}
\usepackage[rel]{overpic}
\graphicspath{{figures/}}

\usepackage{tikz,xcolor,hyperref}
\definecolor{lime}{HTML}{A6CE39}
\DeclareRobustCommand{\orcidicon}{%
	\begin{tikzpicture}
	\draw[lime, fill=lime] (0,0)
	circle [radius=0.16]
	node[white] {{\fontfamily{qag}\selectfont \tiny ID}};
	\draw[white, fill=white] (-0.0625,0.095)
	circle [radius=0.007];
	\end{tikzpicture}
	\hspace{-2mm}
}
\foreach \x in {A, ..., Z}{%
	\expandafter\xdef\csname orcid\x\endcsname{\noexpand\href{https://orcid.org/\csname orcidauthor\x\endcsname}{\noexpand\orcidicon}}
}

\usepackage{graphicx}
\usepackage{newtxtext}
\usepackage{newtxmath}
\usepackage{natbib}
\usepackage{hyperref}

\newcommand{\RomanNumeralCaps}[1]
\linenumbers

\title{Turbulence modulation in particle-laden channel flow: the particle inertial effects}

\author{Zi-Mo Liao\aff{1}\orcidZ{},
Feng-Hui Lin\aff{1}\orcidF{},
Luoqin Liu\aff{1}\orcidL{},
Nan-Sheng Liu\aff{1}\orcidN{}\corresp{\email{lns@ustc.edu.cn}},
Xi-Yun Lu\aff{1}\orcidX{}\corresp{\email{xlu@ustc.edu.cn}}}

\affiliation{\aff{1}Department of Modern Mechanics, University of Science and Technology of China, Hefei 230027, PR China}

\begin{document}
\maketitle

\begin{abstract}
        The particle inertial effects on turbulence modulation in particle-laden channel flow are investigated through four-way coupled point-particle direct numerical simulations, with the mass loading fixed at $0.6$ and friction Stokes number $St^+$ varying from $3$ to $300$.
        A full transition pathway is realized in sequence from a drag-enhanced to a drag-reduced flow regime, before asymptotically approaching the single-phase state as $St^+$ increases continuously up to 300. 
        For the first time, a set of transport equations for the particle phase is derived analytically to interpret the inter-phase coupling, in the context of the point-based statistical description of particle-laden turbulence.
        By virtue of this, two dominant mechanisms are substantially identified and quantified: a positive, particle-induced extra transport, which decreases monotonically with $St^+$, and a negative, particle-induced extra dissipation, which depends non-monotonically on $St^+$.
        The coupling of these two mechanisms leads to a direct contribution of particle phase to the shear stress balance, turbulent kinetic energy, and Reynolds stress budgets.
        As a consequence, with the increase of particle inertia, the self-sustaining cycle of near-wall turbulence transitions from being augmented to being suppressed and, eventually, recovers to the single-phase situation.
        This gives rise to an indirect effect, manifested by the non-monotonic modification of Reynolds shear stress and turbulent production rate.
        Taken together, comprehensive interplays between particle-modified turbulent transport, particle-induced extra transport and dissipation are analyzed and summarized, providing a holistic physical picture composed of consistent interpretations of turbulence modulation.
\end{abstract}




\section{Introduction}
\label{sec:introduction}

Due to its prevalence and importance in natural and industrial processes, particle-laden turbulence (PLT) has garnered 
renewed attention in fluid dynamics and geophysics.
It plays a fundamental role in a wide range of applications, including predicting the transport rates of windblown sand \citep{Tholen2023}, understanding the impact of turbidity currents on global geochemical cycling \citep{Salinas2020,Salinas2021}, and assessing the hazard posed by pyroclastic flows throughout volcanic eruptions \citep{Lube2020}.
A common challenge across these diverse contexts is to unravel the transport dynamics of mass, momentum, and energy in PLT, especially in wall-bounded geometries.
In that, the physical understanding of these intricate transport processes is crucial for both academic interests and practical applications, particularly with regard to particle-rendered turbulence modulation and the consequent flow resistance.
However, the apparently discrepant phenomena reported for the PLT have perplexed the community for decades, for which no compatible interpretations have been proposed in flow physics.
Specifically, a long-standing debate exists regarding whether inertial particles are introduced to augment or suppress the turbulence dynamics and in turn to result in drag enhancement or drag reduction \citep{Rossetti1972, Zhao2010, Costa2021, Gualtieri2023, Cui2024}.

The subtle potential of inertial particles to modify the drag force in wall-bounded turbulent flows has been demonstrated by the pioneering experiment of \citet{Rossetti1972}.
Drag reduction is reported in both vertical and horizontal pipe flows, which is found sensitive to the control parameters including the material properties and the mean concentration of particles.
Similar conclusions have been drawn by subsequent numerical studies in the channel flow configuration \citep{Zhao2010, Zhao2013, Capecelatro2018, Gao2023}.
This particle-induced drag reduction is generally accompanied by a drastic laminarization manifested by the near-wall flow structures, namely, the low-speed streaks are elongated with their spanwise spacings increased.
According to these works, the physical cause of flow laminarization and drag reduction is attributed to the particle(-induced) dissipation, arising due to the non-conservative interphase momentum coupling -- the Stokes drag force \citep{Kulick1994}.
Via the two-way coupled point-particle direct numerical simulations (PP-DNS), \citet{Zhao2013} affirmed that particle dissipation leads to significant kinetic energy loss and damps the velocity fluctuations in wall-normal and spanwise directions.
They, therefore, claimed that it is the net energy loss that results in profound drag reduction behaviours.
On the contrary, other numerical results showed that introducing inertial particles will increase the frictional drag \citep{Costa2021, Gualtieri2023}.
By examining the shear stress balance and turbulent kinetic energy (TKE) budget, \citet{Gualtieri2023} pointed out that additional momentum and energy fluxes induced by the particle phase act as extra shear stress and TKE production, respectively, accounting for the observed drag enhancement.

Despite the above inconsistent findings, the following important studies pointed to a compatible interpretation that rationales both drag reduction and drag enhancement depending on the key parameters.
When focusing on the dilute suspension of small heavy particles, where the volume fraction is negligibly small $\phi_v\ll1$, the density ratio is large $\rho_p/\rho\gg1$, and the particle's diameter $d_p$ is small compared with Kolmogorov length scale $\eta$, two parameters of the particle phase are known to govern the essential flow dynamics of wall-bounded PLT \citep{Bec2024}.
The first is the friction Stokes number defined as $St^+=\tau/t_v$, with $\tau$ being the particle relaxation time and $t_v$ the viscous time scale.
The second is the mass loading defined as $\Phi_m\equiv\phi_v\rho_p/\rho$.
By exploring the particle inertia effects quantified by $St^+$, \citet{Lee2015} showed that turbulence modification depends non-monotonically on $St^+$. 
They found that low-inertia particles ($St^+=0.5$) enhance near-wall turbulence, while moderate-to-high inertia particles ($St^+=5, 35, 100$) have an opposite impact, with the most pronounced suppression occurring at $St^+=35$.
Such a non-monotonic dependence on $St^+$ was also reported for the Reynolds stress budgets by \citet{Dritselis2016}.

On the other hand, it is demonstrated by \citet{Dave2023} that particles of $St^+=6$ are introduced to enhance the flow resistance, while those of $St^+=30$ reduce it. 
Of creative importance, they took use of the conceptions well-established to characterize the drag modification (reduction or enhancement) mechanisms in turbulent polymeric flows \citep{White2008}. 
As a mechanistic explanation, the polymer-induced drag modification has been specified as a result of the reduction in Reynolds stress compensated by the development of polymer stress \citep{Zhu2023,Song2023,Lin2024}.  
In that, \citet{Dave2023} interpreted the above opposite turbulence modulation as a $St^+$-dependent competition between the increased particle shear stress and the reduced Reynolds shear stress. This servers indeed to open a novel path leading to possibly consistent interpretations for different particle-induced drag responses. 
Following this path, a key aspect in fully understanding turbulence modulation in wall-bounded PLT lies in characterizing the underlying mechanisms and their $St^+$ dependencies via proper analytical approaches.

It has been directed that utilizing a statistical approach works to present a viable solution.
In recent years, a point-based statistical description for PLT has been gradually developed, integrating the turbulence theory for the fluid phase with the kinetic theory for the particle phase \citep{Subramaniam2020, Subramaniam2022}.
\citet{Johnson2020} serves as an exemplary demonstration of this theoretical framework, specifically focusing on turbophoresis in one-way coupled channel flows. 
From the particle-phase wall-normal momentum transport equation, an analytical formula, termed the `turbophoresis integral', was derived to relate the particle concentration profile to the velocity statistics explicitly.
Utilizing the turbophoresis integral, both the turbophoresis pseudo-force and the biased-sampling force can be clearly quantified \citep{Soldati2009}, and this approach can also incorporate additional effects, such as the electrostatic drift \citep{Zhang2023Zheng}.
Moreover, as the limit behaviours of the turbophoresis integral are also examined by \citet{Johnson2020}, the integral has been utilized in modeling the particle statistics. 
This approach significantly outperforms previous closure models, particularly in terms of accurately modeling the fluid velocity sampled by particles \citep{Bragg2012pre,Zhang2023}.
In short, the advantages of statistical description in understanding and modeling the near-wall accumulation of small particles have been well demonstrated.

However, the great potential of point-based statistical descriptions in capturing particle-induced turbulence modulation is not exploited well by the inspiring work of \citet{Johnson2020}.  
In contrast to the transport equations for the fluid phase which have been extensively employed to investigate various mechanisms of turbulence modulation \citep{Lee2015,Dritselis2016,Pan2020}, those for the particle phase are still lacking. 
For example, \citet{Johnson2020} and subsequent studies have attempted to examine the particle-phase wall-normal momentum equation to address wall-normal migration, momentum equations in other directions, as well as the kinetic energy equation, remain unexplored yet.
Additionally, the coupling between the particle- and fluid-phase transports, has not yet been addressed.
This gap hinders the proper characterization of interphase coupling mechanisms, leading to often incompatible interpretations or incomplete physical pictures, as discussed above.
Therefore, a comprehensive examination of particle-phase transport processes in the statistical context is essential to provide a consistent explanation of turbulence modulation, and to offer a definitive solution to the long-standing debate on particle-modified flow resistance.
It servers as one of the key motivations of the present work.

In the present study, turbulence modulation induced by small, heavy particles is investigated through four-way coupled PP-DNS of incompressible particle-laden channel flows.
A flow transition from drag-reduced to drag-enhanced states is realized due to the increase of particle inertia (characterized by $St^+$).
To this end, a comprehensive statistical description of the particle-laden channel turbulence is developed to illustrate the inter-phase coupling. 
Specifically, a series of transport equations for momentum, kinetic energy, and Reynolds stresses for the particle phase are derived analytically for the first time (Appendix~\ref{app:notes}).
These equations work to identify and quantify the coupling mechanisms by which the dispersed particles modify the carrying flow, offering a harmonious explanation for the intriguing $St^+$-dependence of particle-induced drag behaviours. 
By virtue of this, a novel physical picture of turbulence modulation has been drawn up, encompassing the flow resistance, the TKE production/dissipation, as well as the self-sustaining processes. 

\section{Mathematical descriptions}
\label{sec:governing-equations}

\subsection{The deterministic description}
\label{sec:governing-equations-deterministic}

Since we are considering a dilute suspension of small heavy particles ($\phi_v\ll1$, $\rho_p/\rho\gg1$, and $d_p<\eta$), the continuous fluid motion is governed by the incompressible Navier-Stokes equations
\begin{align}
\fr{\pat u_i}{\pat x_i}= & \,0, \label{eqn:flow-continuity} \\
\frac{\partial u_{i}}{\partial t}+u_{j}\frac{\partial u_{i}}{\partial x_j}= & -\frac{1}{\rho}\frac{\partial p}{\partial x_i}+\nu\frac{\partial^2 u_i}{\partial x_j\partial x_j}+{\mathcal{F}_i}, \label{eqn:flow-momentum}
\end{align}
where $\pu=(u_i), \,i=1,2,3$ denotes the fluid velocity; $p$ represents the pressure, $\rho$ the density and $\nu=\mu/\rho$ the kinematic viscosity with $\mu$ being the dynamic viscosity of the fluid phase.
The particle feedback force $\mathcal{F}_i$ appearing in the right-hand side of Eq.~(\ref{eqn:flow-momentum}) can be formally written as a summation over all $N_p$ particles \citep{Crowe2011},
\begin{equation}
	\mathcal{F}_i= -\frac{1}{\rho}\sum_{n=1}^{N_p}\delta(\px-\px_p^{(n)})F_i^{(n)}. \label{eqn:feedback}
\end{equation}
Here, the particle indexed by superscript ${(n)}$ has the velocity $\pv=(v_i),\,i=1,2,3$, and the force acting on it through interphase momentum coupling is denoted by $F_i$.
The Dirac delta function $\delta(\,\cdot\,)$, which is approximated numerically by a non-singular mollification kernel \citep{Capecelatro2013}, maps the feedback located at point-particle position $\px_p(t)$ to the Eulerian field.

The dispersed particle phase is described in a Lagrangian perspective.
Considering the monodispersed situation where all physical properties of particles (e.g. the density $\rho_p$, the diameter $d_p$, the mass $m_p=\rho_p\pi d_p^3/6$) are identical, the temporal evolution of each point particle (the particle indices are neglected here for simplicity) is governed by \citep{Crowe2011}
\begin{align}
\frac{\mathrm{d}\px_{p}}{\mathrm{d}t}= & \ \pv, \label{eqn:particle-position} \\ 
m_p\frac{\mathrm{d}\pv}{\mathrm{d}t}=  & \ \pF=3\pi\mu d_p\left[\pu(\px_p,t)-\pv\right]f_D(Re_p). \label{eqn:particle-momentum}
\end{align}
Note that for small heavy particles of interest, interphase momentum couplings except the Stokes drag are negligible \citep{Bec2024}.
The finite Reynolds number effect can be taken into account by introducing the widely adopted Schiller-Naumann correlation $f_D=1+0.15 Re_p^{0.687}$ \citep{Schiller1933}, with particle Reynolds number defined as $Re_p\equiv {\rho|\pu-\pv|d_p}/{\mu}$.
By introducing the particle relaxation time $\tau\equiv\rho_pd_p^2/(18\mu)$ and the slip velocity $\pps=\pu-\pv=(s_i),\,i=1,2,3$, Eq.~\er{eqn:particle-momentum} can be rewritten as 
\beq
\fr{ {\mathrm{d}v_i} }{ {\mathrm{d}t} } = \fr{F_i}{m_p} = \fr{f_D s_i}{\tau}.
\eeq 

The inter-particle collisions can be described using the hard-sphere model \citep[HSM,][]{Hoomans1996,Sundaram1997,Mallouppas2013} which is well suited to dilute suspension of rigid spheres considered here.
Specifically, instead of resolving the interactions during particle-particle contacting, HSM directly computes the results of binary collisions through
\begin{align}
	\pv^{(n)} = & \, \pv^{(n)}_0-\frac{1+e}{2}\left[\pxi\cdot\left(\pv^{(n)}_0-\pv^{(m)}_0\right)\right]\pxi, \label{eqn:hsm-1} \\
	\pv^{(m)} = & \, \pv^{(m)}_0+\frac{1+e}{2}\left[\pxi\cdot\left(\pv^{(n)}_0-\pv^{(m)}_0\right)\right]\pxi, \label{eqn:hsm-2}
\end{align}
where $\pv^{(n)}$ and $\pv^{(m)}$ denote the post-collision velocities and $\pv^{(n)}_0$ and $\pv^{(m)}_0$ the pre-collision velocities;
$\pxi$ is the unit vector pointing from the center of particle $n$ to the center of particle $m$; and $e$ denotes the coefficient of restitution.
In the present simulations, both particle-particle and particle-wall collisions are assumed to be elastic ($e=1$) to exclude the collisional dissipation.

Based on the above equations, we obtain a closed, deterministic description of particle-laden flows.
Integrating the Lagrangian particle tracking (\ref{eqn:particle-position})-(\ref{eqn:particle-momentum}), particle back-reaction (\ref{eqn:feedback}), and inter-particle collision  (\ref{eqn:hsm-1})-(\ref{eqn:hsm-2}), this description serves as the basis of four-way coupled PP-DNS.
An early classification proposed by \citet{Elghobashi1991, Elghobashi1994} conjectured that the inter-particle collision plays a role only when the mean volume fraction $\phi_v$ is above $10^{-3}$.
However, the occurrence of preferential concentration and turbophoresis can result in a drastic increase in local volume fraction, where the inter-particle collisions take into play and thus the four-way coupling becomes indispensable \citep{Subramaniam2022}.

\subsection{The statistical description}
\label{sec:governing-equations-statistical}

For turbulent flows laden with a large number of small heavy particles, it has been established that the statistical description is advantageous for theoretical analysis on the collective behaviours of the particle ensemble \citep{Reeks1991,Reeks1992,Bragg2012pre,Johnson2020}.
A point-based statistical description of particle-laden turbulent flows integrates turbulence theory for the fluid phase with kinetic theory for the particle phase \citep{Subramaniam2020, Subramaniam2022}.
Since all particles are considered identical, their statistical behaviours are characterized by the one-particle probability distribution $f$ in the six-dimensional phase space $(\px,\pv)$ with the normalization condition $\int f\,\mathrm{d}\px\mathrm{d}\pv=1$ (see Appendix~\ref{app:notes-distribution} for a formal definition).
Without any assumption and modeling of phase-space transports, the probability distribution evolution is governed by the following equation \citep{Lifschitz1983},
\begin{equation}
\frac{\partial f}{\partial t}+\frac{\partial (v_k f)}{\partial x_k}+\frac{1}{m_p}\frac{\partial (F_k f)}{\partial v_k}=\left(\frac{\partial f}{\partial t}\right)_\mathrm{coll}. \label{eqn:Boltzmann}
\end{equation}
The right-hand-side term $(\partial f/\partial t)_\mathrm{coll}$, known as the collision integral, represents the contribution of inter-particle collisions on the phase-space evolution.
Here, the force acting on particles, $F_k$, depends on velocity through the Stokes drag and is thus dissipative.
In this work, Eq.~(\ref{eqn:Boltzmann}) will not be handled directly.
Instead, we deal with the derived physical-space transport equations, which also referred to as the moment equations.

As well known, after the particles are released into a fully developed turbulent channel flow, the two-phase system will gradually approach a statistically stationary state.
Furthermore, the intrinsic property of single-phase channel turbulence, namely, statistical homogeneity in the streamwise and the spanwise directions, is preserved, as verified by numerous studies \citep{Marchioli2008,Soldati2009}.
Hence, the mean particle concentration can be formally defined as \citep{Johnson2020}
\begin{equation}
c(y,t) = \idotsint f(\px,\pv;t)\,\mathrm{d} \pv \mathrm{d}x \mathrm{d}z. \label{eqn:particle-concentration}
\end{equation}
It is also referred to as the wall-normal probability density which inherits the normalization condition from $f$.
To effectively characterize the near wall accumulation of particles, it is common to consider the normalized particle concentration $\hat{c}\equiv c/c_b$ where the bulk concentration is $c_b = \int_0^{L_y} c {\rm d} y / L_y = 1/L_y$ and $L_y$ denotes the domain size in $y$ direction.

Moreover, the one-dimensional particle average of a given function $A(\px,\pv)$ is defined as
\begin{equation}
\langle A\rangle(y,t) \equiv \fr{1}{c(y,t)} \idotsint A f\,\mathrm{d}\pv \mathrm{d}x \mathrm{d}z. \label{eqn:particle-average}
\end{equation}
Clearly, setting $A=1$ in Eq.~\er{eqn:particle-average} recovers Eq.~\er{eqn:particle-concentration}.
This definition is consistent with the Reynolds average $\overline{\vphantom{(}\,\cdot\,}$ of flow variables in canonical channel turbulence, which is typically approximated by streamwise and spanwise averages.
It should be noted that the particle average $\langle\,\cdot\,\rangle$ can act not only on particle variables but also on flow variables.
For example, $\langle u_1\rangle$ means the averaged streamwise flow velocity sampled at particle positions, therefore in conception different from its Reynolds averaged counterpart $\bar{u}_1$ due to the preferential sampling nature of inertial particles \citep{Bragg2012pre,Capecelatro2016}.
Unlike the Reynolds decomposition ($A=\bar{A}+A'$), the fluctuating part of variable $A$ with regard to its particle average is denoted by $A^\circ\equiv A-\langle A\rangle$ hereafter.

Following a procedure analogous to that of deriving the Navier-Stokes equations from the Boltzmann equation \citep{Kremer2010}, the averaged transport equations for the particle phase can be easily obtained.
Once the two-phase system approaches the statistically stationary state where ${\partial\langle\,\cdot\,\rangle}/{\partial t}={\partial \overline{\vphantom{(}\cdot\vphantom{)}}}/{\partial t}=0$, the particle-phase continuity equation leads to $\langle v_i\rangle=(\langle v_1\rangle(y),\,0,\,0)$; and the corresponding momentum equation is formulated as (Appendix~\ref{app:stationary})
\begin{equation}\label{eqn:pmt}
\frac{\mathrm{d} (C\langle v_iv_2\rangle)}{\mathrm{d} y} - {C\langle\widehat{F_i}\rangle} = 0. 
\end{equation}
Here we define $\widehat{F}_i\equiv F_i/m_p$ and \beq\lb{eq.C}
C(y,t) = \Phi_m \hat{c} = \Phi_m L_y \idotsint f(\px,\pv;t)\,\mathrm{d} \pv \mathrm{d}x \mathrm{d}z,
\eeq 
so that Eq.~(\ref{eqn:pmt}) represents the conservation of particle-phase momentum per unit mass, namely, a balance realized between particle transport (the first term) and fluid driving force (the second term).
The weight function $C(y,t)$ reflects the compressible nature of the particle phase, which manifests in wall turbulence as local accumulation due to turbophoresis \citep{Fox2014}.
Furthermore, the above equation can be inherently connected with the Reynolds averaged momentum equation of the fluid phase through (Appendix~\ref{app:coupling})
\begin{equation}
\overline{\mathcal{F}}_i=-C\langle \widehat{F_i} \rangle. \label{eqn:coupling}
\end{equation}
Such a procedure can also be applied to the second-order particle velocity moments, similar to that of deriving the TKE ($k=\overline{u'_iu'_i}/2$) and Reynolds stress ($r_{ij}=\overline{u'_iu'_j}$) budgets, leading to the budget equations of particle-phase fluctuating kinetic energy (PFKE), $k_p \equiv \langle v_i^\circ v_i^\circ\rangle/2$, and particle-phase Reynolds stresses, $r_{p,ij}\equiv \langle v_i^\circ v_j^\circ\rangle$.
The derivations of these transport equations and the coupling relations between the fluid and particle phases are detailed in Appendices~\ref{app:derivation}-\ref{app:coupling}.
The above statistical methodology provides a theoretical tool of significance, as demonstrated thoroughly below in \S~\ref{sec:results}, where the particle-phase transport dynamics are analysed in detail to fulfill the physical understanding of particle-laden turbulent flows.

\section{Dataset descriptions}
\label{sec:dataset}

In the present study, near-incompressible particle-laden channel flows are solved using the GPU-accelerated four-way PP-DNS code we developed recently for the compressible PLT problems \citep{Liao2024}. To minimize the compressibility effects, a low Mach number of $0.2$ is considered according to \citet{Gerolymos2014} and \citet{Modesti2016}, corresponding to a friction Mach number as small as $0.011$; commensurately, the mean variation and maximum root-mean-square of temperature profiles are obtained negligibly small, i.e. $(T_c-T_w)/T_w=0.0076$ and $\max{T_\text{rms}(y)}/T_w=0.0010$, where the subscripts $c$ and $w$ are used to denote variables at channel centerline and the wall, respectively.
The numerical solver has been extensively validated by accurately reproducing the statistical results of several benchmark simulations of incompressible particle-laden channel flows \citep{Marchioli2008, Kuerten2011, Johnson2020, Dave2023}, encompassing all three interphase coupling levels (namely, the one-, two-, and four-way couplings), see \citet{Liao2024} and Appendix~\ref{app:validation} for code validation.

Here we briefly introduce the numerical methods adopted in this GPU-based solver of high parallel efficiency.
The governing equations of the fluid phase are solved on a rectilinear grid using a high-order finite-difference method. Specifically, the convective and viscous terms are discretized using an 8th-order energy-preserving scheme \citep{Pirozzoli2010} and a 6th-order central scheme, respectively.
A three-stage, third-order Runge-Kutta (RK) scheme is used for time integration \citep{Spalart1991, Bernardini2021}.
The flow variables seen by particles are computed by trilinear interpolation;
the approximation of particle feedback (\ref{eqn:feedback}) is implemented using trilinear extrapolation;
and the collision detection is achieved using the spatial subdivision technique.
The third-order total variation diminishing RK is applied to advance particle variables in time \citep{Gottlieb1998}.
Further details on the numerical techniques and GPU implementation strategies can be found in \citet{Liao2024}.

\begin{figure}
    \centering
    \includegraphics[width=0.8\linewidth]{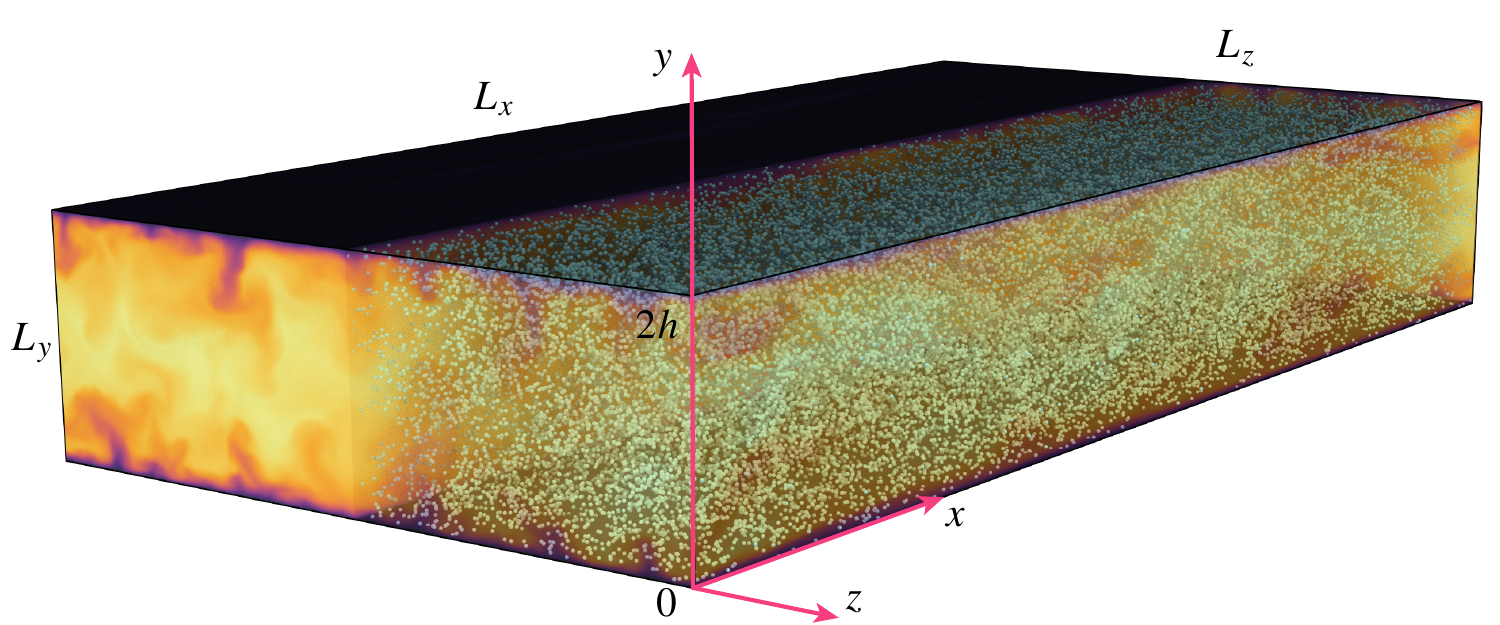}
    \caption{Schematic of the computational domains used in the present simulations. The contour plot represents an instantaneous streamwise velocity field for the St030 case. Half of the computational domain is set to be translucent to visualize the dispersed particles (light blue spheres).}
    \label{fig:configuration}
\end{figure}

For all the simulations conducted, the bulk Reynolds number $Re_b\equiv U_b h/\nu=5150$ is fixed to ensure a constant flow rate and, correspondingly, the friction Reynolds number $Re_\tau\equiv u_\tau \delta_v/\nu$ is obtained around $300$.
Here $U_b$ denotes the bulk velocity and $h$ the half channel height. 
The wall units are defined as follows: the friction velocity $u_\tau\equiv\sqrt{\tau_w/\rho}$ ($\tau_w$ denotes the wall shear stress), the viscous length scale $\delta_v\equiv\nu/u_\tau$, and the viscous time scale $t_v\equiv\nu/u_\tau^2$.
The computational box size is $L_x\times L_y\times L_z=4\pi h\times 2h\times 2\pi h$, as illustrated in figure~\ref{fig:configuration}.
The grid sizes used in the $x,y,z$ directions are $N_x\times N_y\times N_z=512\times 192\times 384$, corresponding to grid spacings of $\Delta x^+\approx5.6$, $0.4\lesssim\Delta y^+\lesssim5.6$, $\Delta z^+\approx5.0$. 
As a common treatment, the rectilinear grid is stretched in the wall-normal direction to ensure a sufficient near-wall grid resolution for DNS.
Hereafter, variables with superscript `$+$' are nondimensionalized by the wall units. 

Initially, inertial particles are randomly released into a fully developed single-phase channel flow, with velocities interpolated by the surrounding fluid.
The streamwise and spanwise boundary conditions for particles are assumed periodic to maintain a constant mass loading.
The parameter set-ups for the particle phase of all cases are listed in table~\ref{tab:param-part}, having a wide Stokes number range of two orders, i.e. $3\leq St^+\leq300$.
The particle diameter is fixed at $d_p^+=d_p/\delta_v^\text{SP}=0.5$ and thus the friction Stokes number $St^+\equiv\tau/t_v^\text{SP}=(\rho_p/\rho)(d_p/\delta_v^\text{SP})^2/18$ is determined by density ratio $\rho_p/\rho$. Hereafter the abbreviation $\text{SP}$ denotes the variables or the situation for the single-phase flow.
Following \citet{Zhou2020} and \citet{Dave2023}, the particle mass loading ${\Phi}_m$ is fixed at $0.6$ for all cases to ensure a considerable turbulence modulation.
After the flow system reaches a statistically stationary state, ensemble averaging is performed over $7500$–$18000$ snapshots to yield flow and particle statistics, corresponding to a sampling window size of $\mathcal{T}/t_v\sim O(10^4)$. 
Additionally, the particle concentration and other variables are sampled and averaged on a wall-normal stretched grid, using the same discretization as the fluid-phase solver, i.e., the slabs adopted for computing particle-phase statistics are centered around each grid point.

\begin{table}
\setlength{\tabcolsep}{10pt}
\centering
\begin{tabular}{cccccccccccc}
Case  & $d_p^+$ & $St^+$ & $N_p$ & $\phi_v$ & ${\Phi}_m$ & $\rho_p/\rho$ \vspace*{2ex} \\
St003 & $0.5$   & $3$    & $1.89\times10^8$ & $2.76\times10^{-3}$   & $0.6$ & $2.16\times10^2$ \\
St010 & $0.5$   & $10$   & $5.67\times10^7$ & $8.27\times10^{-4}$   & $0.6$ & $7.20\times10^2$ \\
St030 & $0.5$   & $30$   & $1.89\times10^7$ & $2.76\times10^{-4}$   & $0.6$ & $2.16\times10^3$ \\
St056 & $0.5$   & $56$   & $1.06\times10^7$ & $1.55\times10^{-4}$   & $0.6$  & $4.03\times10^3$ \\
St100 & $0.5$   & $100$  & $5.67\times10^6$ & $8.27\times10^{-5}$   & $0.6$  & $7.20\times10^3$ \\
St178 & $0.5$   & $178$  & $3.19\times10^6$ & $4.65\times10^{-5}$   & $0.6$  & $1.28\times10^4$ \\
St300 & $0.5$   & $300$  & $1.89\times10^6$ & $2.76\times10^{-5}$   & $0.6$  & $2.16\times10^4$ \\
\end{tabular}
\caption{\label{tab:param-part}
Simulation set-ups for the particle phase. The particle volume fraction $\phi_v$ varies from $\sim10^{-3}$ to $\sim10^{-6}$ corresponding to a mass loading ${\Phi}_m$ fixed at 0.6, lying in the (semi-)dilute suspension regime.}
\end{table}


\section{Results and discussions}
\label{sec:results}

\subsection{The mean flow properties}
\label{sec:mean}

\begin{figure}
\centering
\includegraphics[width=0.96\linewidth]{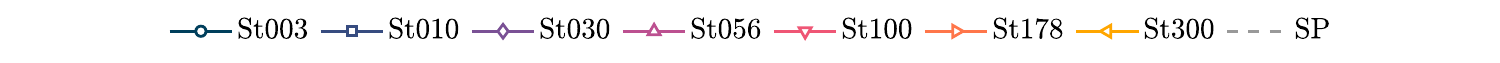}
\begin{overpic}[width=0.32\linewidth]{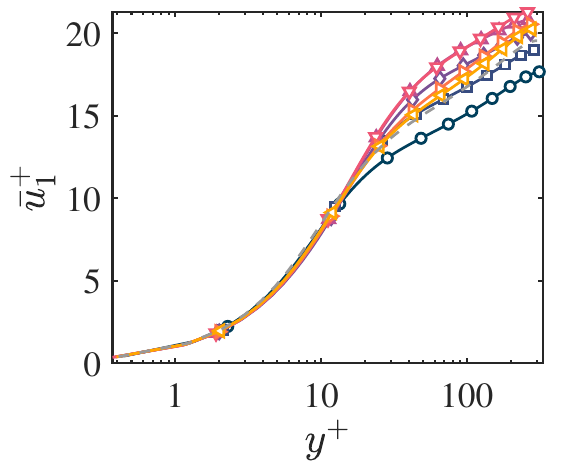}\put(-1,77){$(a)$}\end{overpic}
\begin{overpic}[width=0.32\linewidth]{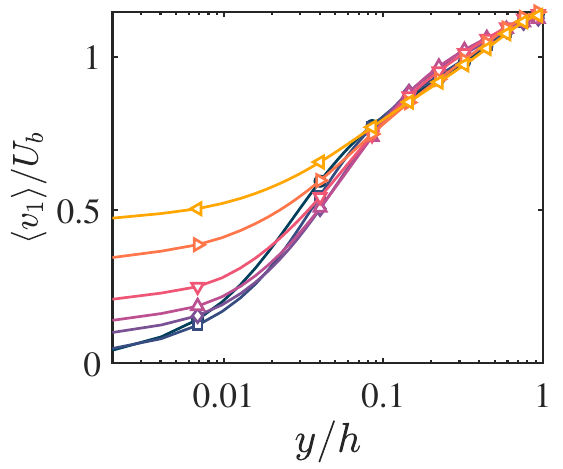}\put(-1,77){$(b)$}\end{overpic}
\begin{overpic}[width=0.32\linewidth]{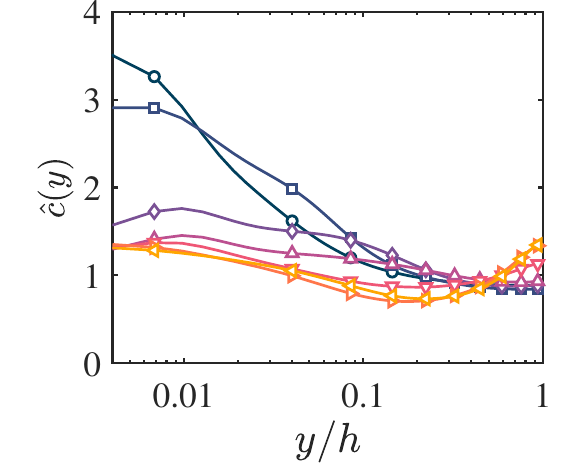}\put(-1,77){$(c)$}\end{overpic}
\caption{\label{fig:mean}The mean profiles of: $(a)$ streamwise flow velocity in wall unit, $(b)$ streamwise particle velocity, and $(c)$ normalized particle concentration. 
}
\end{figure}

At first, we briefly examine the mean flow properties to establish a basic impression of the particle inertial effects on turbulence modulation.
Figure~\ref{fig:mean}$(a)$ depicts the mean flow velocity profiles for all cases listed in table~\ref{tab:param-part}.
Compared with the single-phase results indicated by a dashed gray curve, the linear scaling in the viscous sublayer remains, while the velocity profile around the logarithmic layer is significantly altered.
In the logarithmic layer, the mean velocity is shifted downwards due to the introduction of particles at $St^+=3$.
As the Stokes number increases to $56$, the mean velocity increases successively, exceeding the single-phase one. 
With a further increase of $St^+$ to $300$, the mean velocity decreases and tends to approach the single-phase profile. This variation of the mean flow velocity profile with regard to the Stokes number corresponds to a non-monotonic modification of wall shear stress, i.e., an intriguing flow transition from a drag-enhanced to a drag-reduced regime and subsequently approaching the single-phase state, as substantiated below in the next section.

Given its critical role in momentum and energy transport, the mean particle velocity, $\langle v_1\rangle/U_b$, is also examined for different Stokes numbers in figure~\ref{fig:mean}$(b)$.
It is demonstrated that, as $St^+$ increases, the mean particle velocity profile becomes more and more flattening, especially in the wall region.
This trend has been explained by \citet{Johnson2020ijmf} and \citet{Liao2024} as follows: with the increase of $St^+$, the particle inertia tends to dominate over the Stokes drag exposed by the background flow, leading to an increasingly stronger tendency for the particle phase to approach a so-called many-particle system that has a homogeneous and isotropic equilibrium state.
Moreover, the particle concentration close to the wall obtains a substantial decrease versus $St^+$, see figure~\ref{fig:mean}$(c)$.
It deviates from the pattern observed in one-way coupled simulations, where turbophoresis is found leading to a dramatic increase of near-wall particle concentration over $2\sim3$ orders compared to the uniform distribution and the highest accumulation typically occurring at $St^+\approx30$ \citep{Marchioli2002, Jie2022}.
This deviation can be attributed to the synergistic effect of particle back-reaction and inter-particle collision.
The former alters the intrinsic length and time scales of near-wall coherent structures, which are crucial for the preferential accumulation of inertial particles \citep{Motoori2022, Jie2022}.
The latter induces a strong diffusion of the particle phase, preventing excessive clustering and promoting a more uniform distribution.

\subsection{The drag properties}
\label{sec:drag}

We then focus on the drag properties, which are widely regarded as crucial in wall-bounded turbulence of practical relevance.
The overall statistical results are summarized in table~\ref{tab:statistics}.
It is intriguing to find that the drag modification depends non-monotonically on the Stokes number and the system experiences both drag-enhancement (DE, $St^+\leq10$) and drag-reduction (DR, $St^+\geq30$) regimes.
Specifically, the maximum DE occurs at the lowest $St^+$, with a smooth transition from DE to DR as $St^+$ increases, where $St^+ \approx 10$ marks the critical point.
Further increasing the Stokes number, the flow statistics indicate a tendency to return to the SP ones, resulting in a maximum DR observed at a moderate Stokes number, $St^+\approx100$.

\begin{table}
\centering
\setlength{\tabcolsep}{10pt}
\begin{tabular}{ccccr} 
Case  & $St^+$ & $Re_\tau$ & $C_f(\times10^{-3})$ & DM ($\%$) \vspace*{2ex} \\
St003 & 3   & $334.2$ & $8.375$ & $19.8$  \\
St010 & 10  & $308.7$ & $7.141$ & $2.1$   \\
St030 & 30  & $292.1$ & $6.385$ & $-8.7$  \\
St056 & 56  & $280.5$ & $5.886$ & $-15.8$ \\
St100 & 100 & $277.8$ & $5.774$ & $-17.4$ \\
St178 & 178 & $291.5$ & $6.362$ & $-9.0$  \\
St300 & 300 & $296.8$ & $6.599$ & $-5.6$  \\
SP    & -   & $305.3$ & $6.992$ & $0$   \\
\end{tabular}
\caption{\label{tab:statistics}Overall statistics of the present simulations. Here the skin-friction coefficient is $C_f\equiv2\tau_w/(\rho U_b^2)$. The drag-modification coefficient is defined as $\text{DM}\equiv(C_f-C_f^\text{SP})/C_f^\text{SP}$.}
\end{table}

Such a DE-DR-SP transition can be substantiated by the shear stress balance \citep[SSB, ][]{Costa2021, Dave2023}
\begin{equation}	\underbrace{\vphantom{\int_{0}^{y}}\mu\frac{\mathrm{d}\bar{u}}{\mathrm{d}y}}_{\bar{\tau}^{\text{V}}}
	\underbrace{\vphantom{\int_{0}^{y}}-\rho\overline{u'_1u'_2}}_{\bar{\tau}^{\text{T}}}	+\underbrace{\vphantom{\int_{0}^{y}}\rho\int_{0}^{y}\overline{\mathcal{F}}_1\,\mathrm{d}y}_{\bar{\tau}^{\text{P}}}
	=\tau_w\left(1-\frac{y}{h}\right). \label{eqn:ssb}
\end{equation}
Here $\bar{\tau}^{\text{V}}$ denotes the viscous shear stress of mean flow, $\bar{\tau}^{\text{T}}$ the Reynolds shear stress of turbulent motions, $\bar{\tau}^{\text{P}}$ the extra shear stress induced by particles.
Essentially, these shear stresses reflect the transport of fluid-phase streamwise momentum in the wall-normal direction.
Following \citet{Gualtieri2023}, their contributions to the skin-friction coefficient can be quantified by the well-known FIK-type integrals \citep{Fukagata2002}, i.e. integrating (\ref{eqn:ssb}) twice and reading as
\begin{equation}
\underbrace{\vphantom{\frac{6}{h^2U_b^2}}\frac{6}{Re_b}}_{C_f^{\text{V}}}
+\underbrace{\vphantom{\frac{6}{h^2U_b^2}}\frac{6}{h^2U_b^2}\int_{0}^{h}(h-y)\overline{u'_1u'_2}\,\mathrm{d}y}_{C_f^{\text{T}}}
+\underbrace{\vphantom{\frac{6}{h^2U_b^2}}\frac{6}{h^2U_b^2}\int_{0}^{h}\int_{0}^{y}\int_{0}^{y_1}\overline{\mathcal{F}}_1(y_2)\,\mathrm{d}y_2\mathrm{d}y_1\mathrm{d}y}_{C_f^{\text{P}}}
=C_f.
\end{equation}
Since the bulk Reynolds number is fixed in the present simulations, $C_f^{\text{V}}$ remains constant.
Hence, the frictional drag is determined by an effective shear stress $\bar{\tau}^{\text{E}}$ defined as the summation of $\bar{\tau}^{\text{T}}$ and $\bar{\tau}^{\text{P}}$.

\begin{figure}
\centering
\begin{overpic}[width=0.48\linewidth]{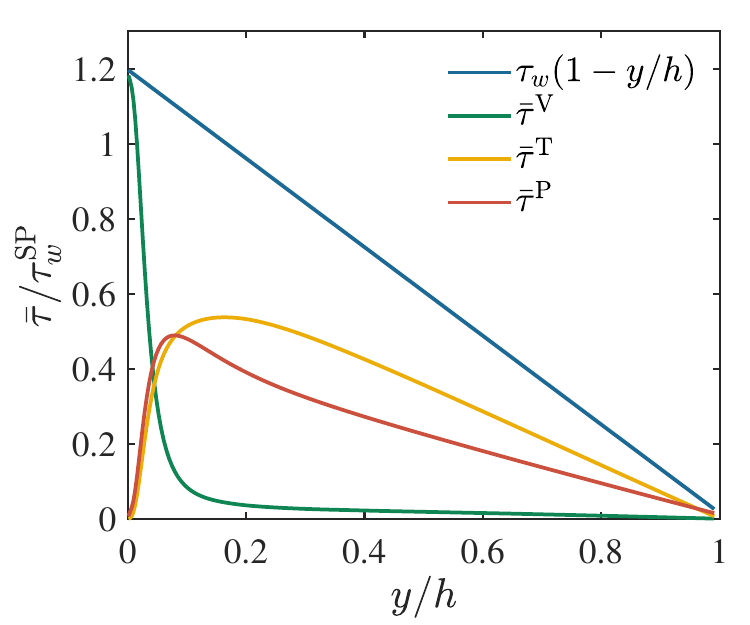}
\put(0,77){$(a)$}
\put(33,72){$St^+=3$}
\end{overpic}
\begin{overpic}[width=0.48\linewidth]{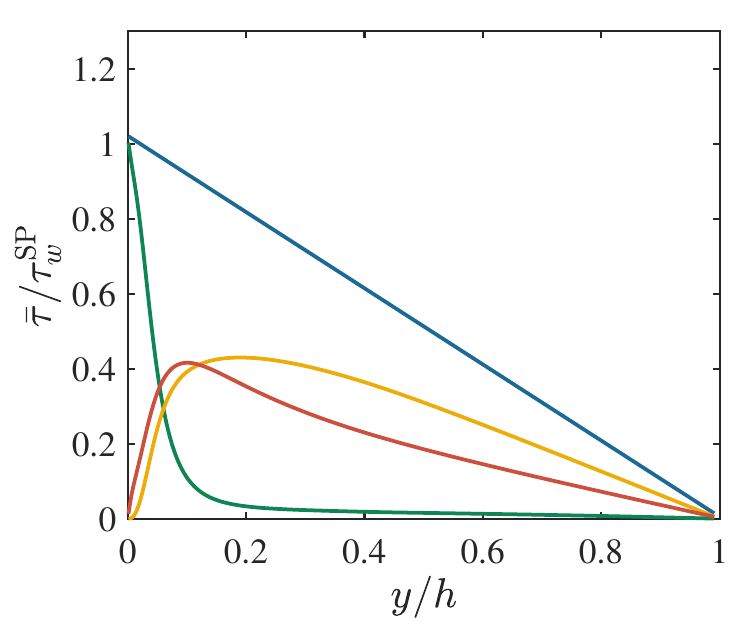}
\put(0,77){$(b)$}
\put(33,72){$St^+=10$}
\end{overpic}
\begin{overpic}[width=0.48\linewidth]{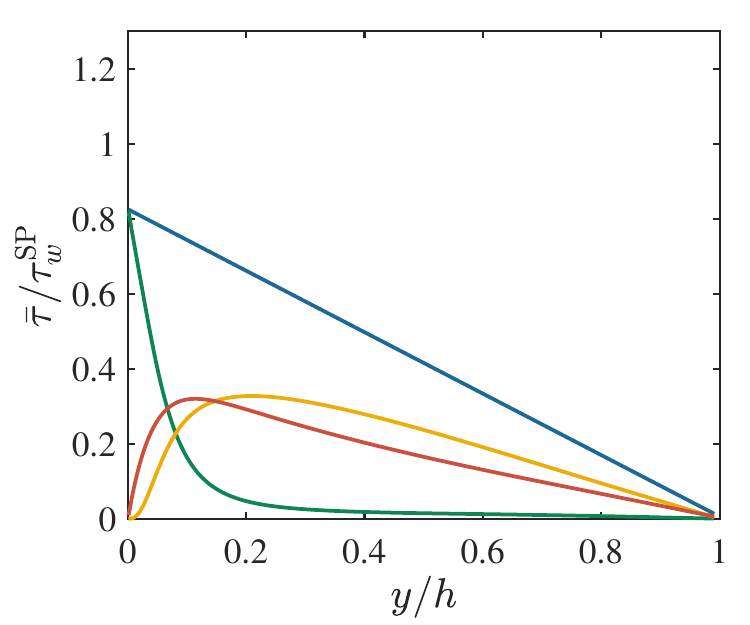}
\put(0,77){$(c)$}
\put(33,72){$St^+=100$}
\end{overpic}
\begin{overpic}[width=0.48\linewidth]{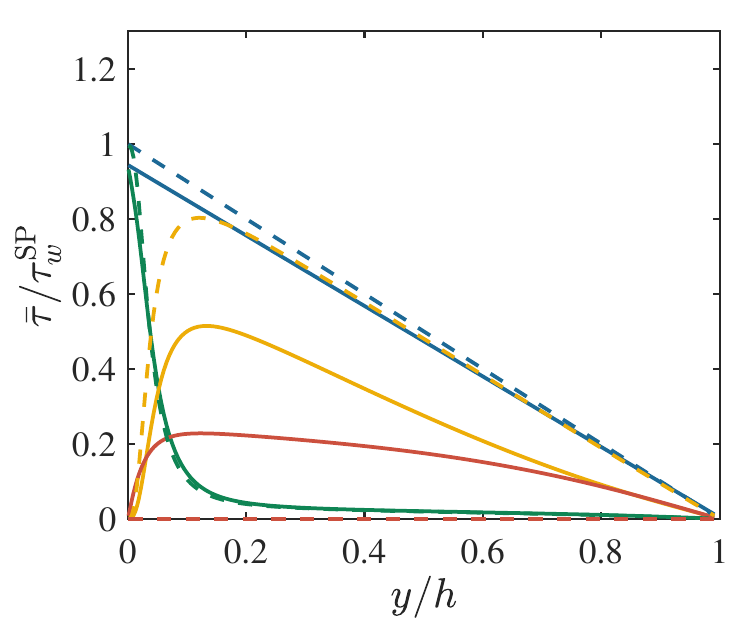}
\put(0,77){$(d)$}
\put(33,72){$St^+=300$}
\end{overpic}
\caption{\label{fig:ssb}The shear stress balance profiles for four typical cases: $(a)$ St003, $(b)$ St010, $(c)$ St100, and $(d)$ St300. In panel $(d)$, the SP results are also depicted in dashed curves with corresponding colors, serving as the $St^+\to\infty$ limit of SSB.}
\end{figure}

\begin{figure}
\centering
\begin{overpic}[width=0.48\linewidth]{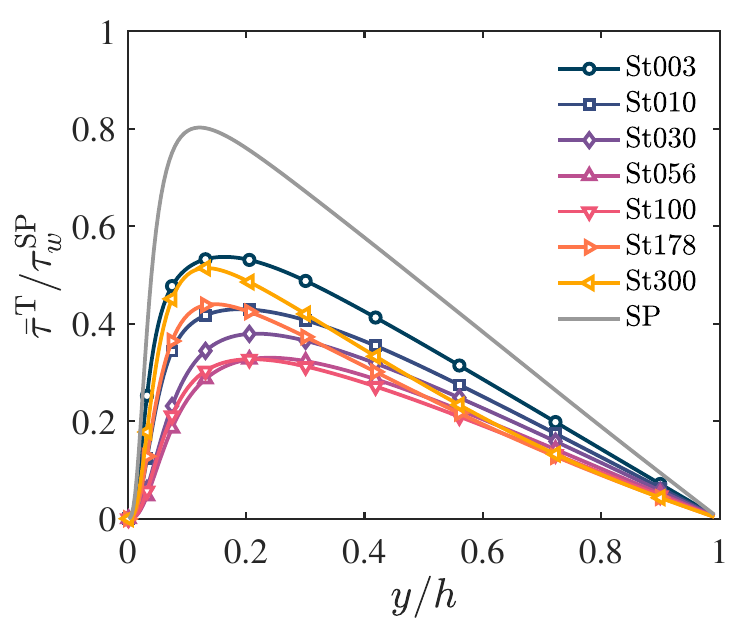}\put(0,77){$(a)$}\end{overpic}
\begin{overpic}[width=0.48\linewidth]{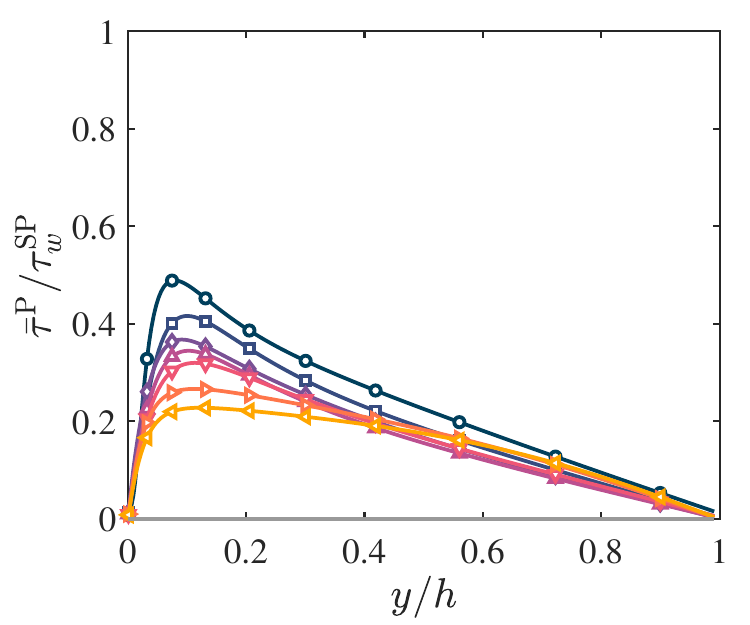}\put(0,77){$(b)$}\end{overpic}
\begin{overpic}[width=0.48\linewidth]{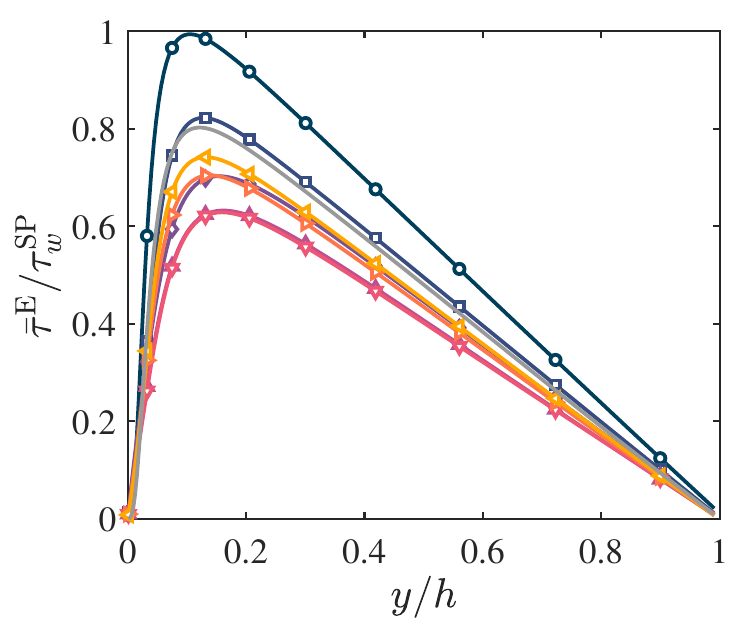}\put(0,77){$(c)$}\end{overpic}
\begin{overpic}[width=0.48\linewidth]{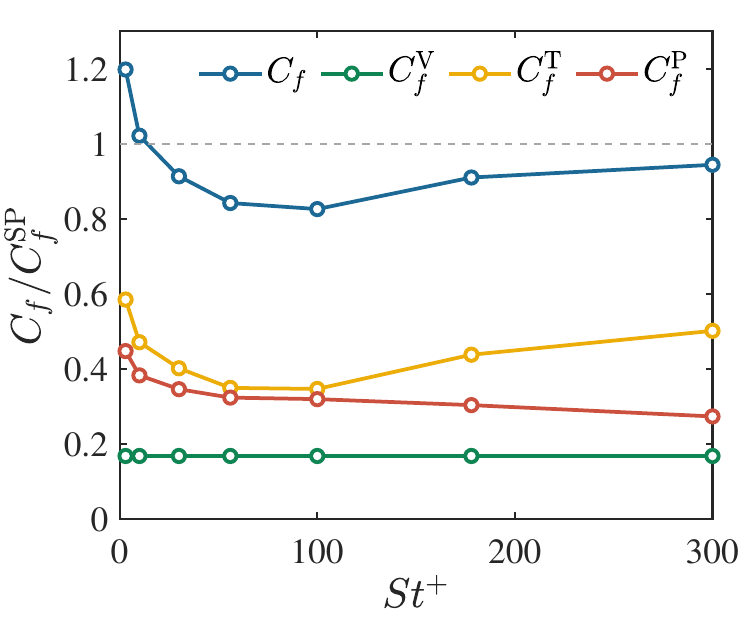}\put(0,77){$(d)$}\end{overpic}
\caption{\label{fig:ssb-st}The profiles of $(a)$ $\bar{\tau}^{\text{T}}$, $(b)$ $\bar{\tau}^{\text{P}}$, and $(c)$ $\bar{\tau}^{\text{E}} = \bar{\tau}^{\text{T}} + \bar{\tau}^{\text{P}}$, along with the $(d)$ $C_f$-$St^+$ curves.}
\end{figure}

We depict four typical SSB profiles (St003, St010, St100, and St300) in figure~\ref{fig:ssb}.
Specifically, in the St003 case, the introduction of $\bar{\tau}^{\text{P}}$ exceeds the suppression of $\bar{\tau}^{\text{T}}$, resulting in a DE of $19.8\%$ (see table~\ref{tab:statistics}).
In the St010 case, $\bar{\tau}^{\text{P}}$ compensates for the suppression of $\bar{\tau}^{\text{T}}$, leading to a flow resistance close to the single-phase case.
In the St100 case, $\bar{\tau}^{\text{P}}$ drops substantially along with a further suppression of $\bar{\tau}^{\text{T}}$, resulting in a DR of $-17.4\%$ (see table~\ref{tab:statistics}).
In the St300 case, the degree of suppression of $\bar{\tau}^{\text{T}}$ is reduced commensurate with a relatively small $\bar{\tau}^{\text{P}}$, causing the flow system to tend closer to the SP situation.
To better illustrate the Stokes number dependence of each term, the $\bar{\tau}^{\text{T}}$, $\bar{\tau}^{\text{P}}$, and $\bar{\tau}^{\text{E}}$ profiles of all cases are shown in figure~\ref{fig:ssb-st}$(a$-$c)$.
As $St^+$ increases, it is demonstrated that basically the Reynolds shear stress decreases as $St^+\leq56$ and then increases when $St^+\geq100$ (figure~\ref{fig:ssb-st}$(a)$), while the particle-induced shear stress has an overall monotonically decreasing trend despite a slight increase of $\bar{\tau}^\text{P}$ for $y/h>0.5$ at high $St^+\geq 100$ (figure~\ref{fig:ssb-st}$(b)$).
Their superposition leads to a non-monotonic variation of the effective shear stress versus $St^+$ (figures~\ref{fig:ssb-st}$(c)$).
Such a competition between the particle-induced extra transport and the suppressed turbulent transport is clearly demonstrated by the $C_f$-$St^+$ curves depicted in figure~\ref{fig:ssb-st}$(d)$.
In short, by exploring the drag modification properties over a wide range of friction Stokes numbers (covering two orders of magnitude), we confirm and significantly extend the competition arguments proposed by \citet{Dave2023}.
A return-to-SP regime is found at high $St^+$, with the critical Stokes numbers for the DE-DR transition and the maximum DR being identified.

\begin{figure}
\centering
\begin{overpic}[width=0.48\linewidth]{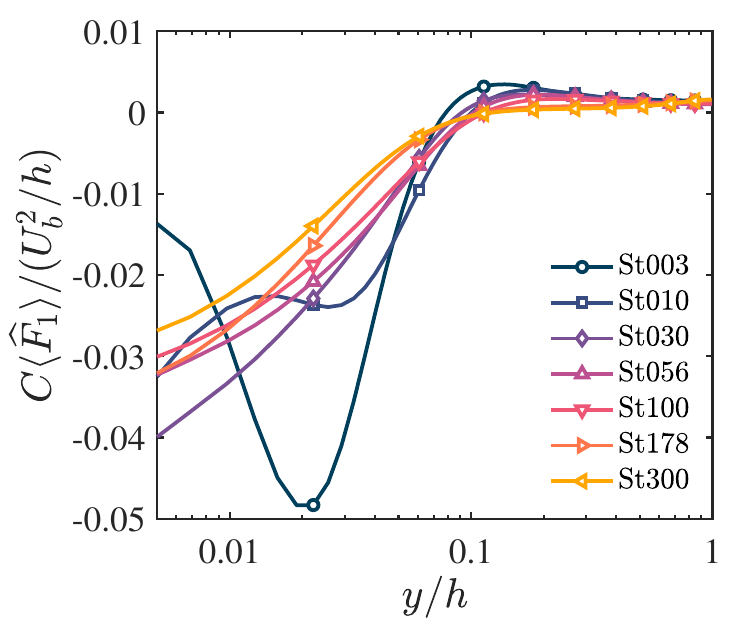}\put(-1,77){$(a)$}\end{overpic}
\begin{overpic}[width=0.48\linewidth]{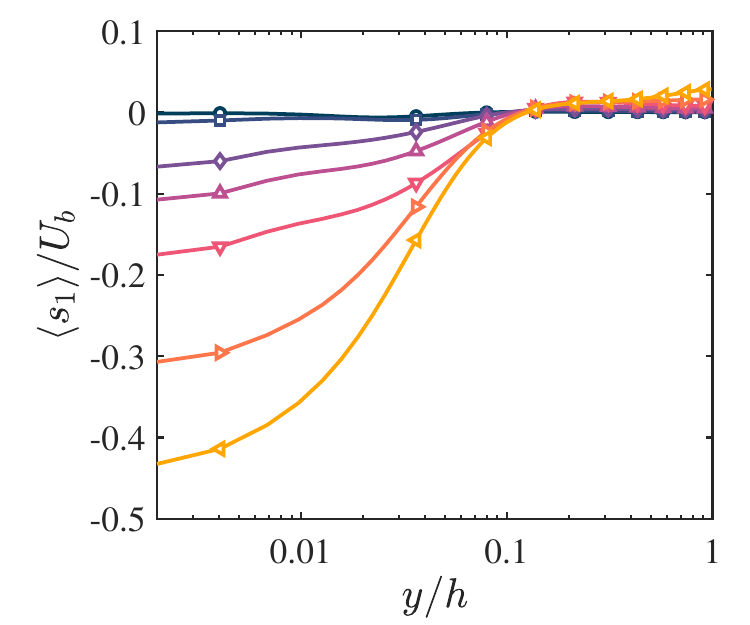}\put(-1,77){$(b)$}\end{overpic}
\caption{\label{fig:pmt}The profiles of $(a)$ streamwise momentum coupling term and $(b)$ mean streamwise slip velocity. 
}
\end{figure}

To gain some new insights into the particle-induced shear stress, we next direct our focus to the particle-phase transport.
The interphase momentum coupling term $C\langle\widehat{F}_1\rangle$ in Eq.~(\ref{eqn:pmt}) is depicted in figure~\ref{fig:pmt}$(a)$, which is the opposite of the particle feedback force $\overline{\mathcal{F}}_1$ according to (\ref{eqn:coupling}).
As shown, in the statistical sense, particles always gain high streamwise momentum from the background flow in the bulk ($C\langle\widehat{F}_1\rangle>0$), transport it toward the wall, and then release high momentum to the near-wall low-speed fluids ($C\langle\widehat{F}_1\rangle<0$).
This bulk-to-wall transport of high momentum undertaken by particles results from the intrinsic wall-normal inhomogeneity of carrying fluid flow and the flattening of particle velocity profile that we have discussed in \S~\ref{sec:mean}.
Given the fact that interphase momentum coupling is proportional to the slip velocity $s_i=u_i-v_i$ shown in figure~\ref{fig:pmt}$(b)$, $C\langle\widehat{F}_1\rangle$ should always be negative near the wall and positive in the bulk, while $\overline{\mathcal{F}}_1$ has just the opposite feature.
Further, since there is no mean streamwise momentum flux at the wall and the channel centerline, i.e. $\langle v_1v_2\rangle|_{y=0,\,h}\equiv0$, integrating Eq.~(\ref{eqn:pmt}) over the channel yields the vanishing of $\int_{0}^{h}C\langle\widehat{F}_1\rangle\,\mathrm{d}y$, as well as $\int_{0}^{h}\overline{\mathcal{F}}_1\,\mathrm{d}y=\bar{\tau}^\text{P}(h)=0$.
Combining the above properties, we have $\bar{\tau}^\text{P}(y)>0$ with $0<y<h$ and $C_f^{\text{P}}>0$, consistent with the numerical results plotted in figure~\ref{fig:ssb-st}$(b,d)$.
In other words, inertial particles provide an additional way to transport high streamwise momentum of fluid flow from the bulk to the wall, in addition to that by turbulent motions, contributing a positive part like the Reynolds shear stress to the frictional drag.

Based on the above analysis, we further attempt to explain why there emerges a DE regime at low Stokes numbers and a return-to-SP regime at high Stokes numbers.
As well known, low-inertia ($St^+\to0$) particles act as flow tracers, while high-inertia ($St^+\to\infty$) particles have a ballistical behaviour \citep{Brandt2022}.
In both situations, due to the mismatch between the time scales of particles and flow structures, particles will not preferentially accumulate in carrying fluids \citep{Motoori2022, Jie2022}.
Theoretically, this results in a uniform distribution of particles, namely, $C\to {\Phi}_m$.
Consequently, the uniformly distributed particles sample the background flow field unbiasedly, degrading the particle-averaged flow variables to their Reynolds-averaged counterparts, i.e., $\langle A_f\rangle\to\overline{A_f}$.
When $St^+\to0$ or $\tau\to0$, particles respond to the fluid flow immediately, thus all the particle statistics reduce to their fluid-phase counterparts.
Equation~(\ref{eqn:pmt}) leads to $\langle\widehat{F}_i\rangle_0=\mathrm{d}(\overline{u'_i u'_2})/\mathrm{d}y$, where $\langle\,\cdot\,\rangle_0$ denotes the particle average at $St^+\to0$ limit.
This means that the particle-induced shear stress formulates as
\beq 
\bar{\tau}^{\text{P}} = \rho \int_{0}^{y} \overline{\mathcal{F}}_1 \,\mathrm{d} y = -\rho C \langle v_1 v_2 \rangle \to -\rho {\Phi}_m \overline{u'_1 u'_2}, \quad St^+\to 0,
\eeq 
and the modified SSB becomes
\begin{equation}
{\mu\frac{\mathrm{d}\bar{u}}{\mathrm{d}y}}
{-\rho(1+{\Phi}_m)\overline{u'_1u'_2}}
=\tau_w\left(1-\frac{y}{h}\right), \quad St^+\to0. \label{eqn:ssb-0}
\end{equation}
It manifests that the effect of particle tracers is to amplify the turbulent transport, namely, the Reynolds shear stress, by a factor of $1+{\Phi}_m$, corresponding to a DE asymptotic state expected for $St^+\to0$.
In contrast, for the other limit of $St^+\to\infty$ or $\tau\to\infty$, all flow properties discussed are indicated to have a tendency to approach the SP case.
It is straightforward to rationalize this from the Stokes drag force formulated as $F_i\approx s_i/\tau$.
Since the slip velocity is certain to be finite, as $\tau$ approaches infinity and ${\Phi}_m$ remains fixed, the feedback force acting on the carrying fluids vanishes \citep{Saffman1962}. 
Hence, the flow modulation by the particle phase should be trivial.
Clearly, these two limits are deemed to bring up bounds for the particle-induced shear stress.
Thus it makes sense that for a finite Stokes number, the positive-valued particle-induced shear stress follows a continuous change from $-\rho{\Phi}_m\overline{u'_1u'_2}$ to zero, leading to the monotonically decreasing $C_f^{\text{P}}$ as observed in figure~\ref{fig:ssb-st}$(d)$.

\subsection{The production and dissipation rates}
\label{sec:tke}

\begin{figure}
\centering
\begin{overpic}[width=0.48\linewidth]{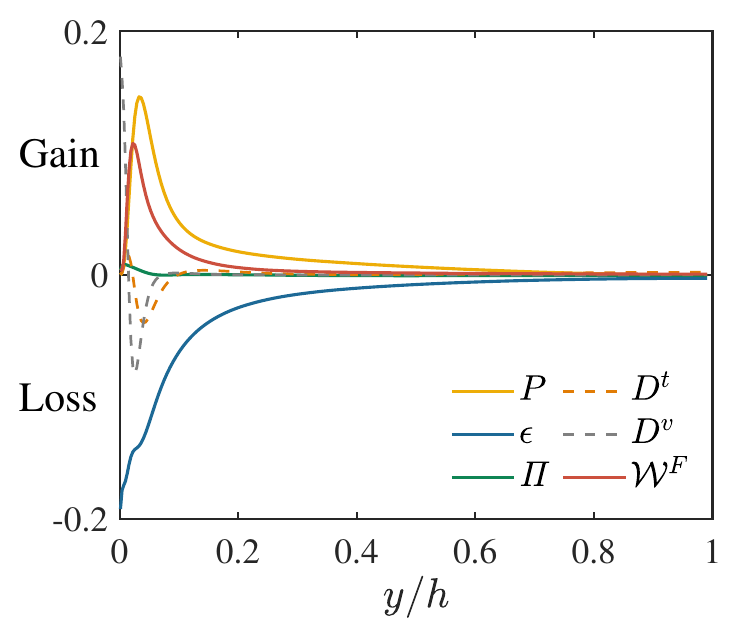}
\put(-1,77){$(a)$}
\put(63,72){$St^+=3$}
\end{overpic}
\begin{overpic}[width=0.48\linewidth]{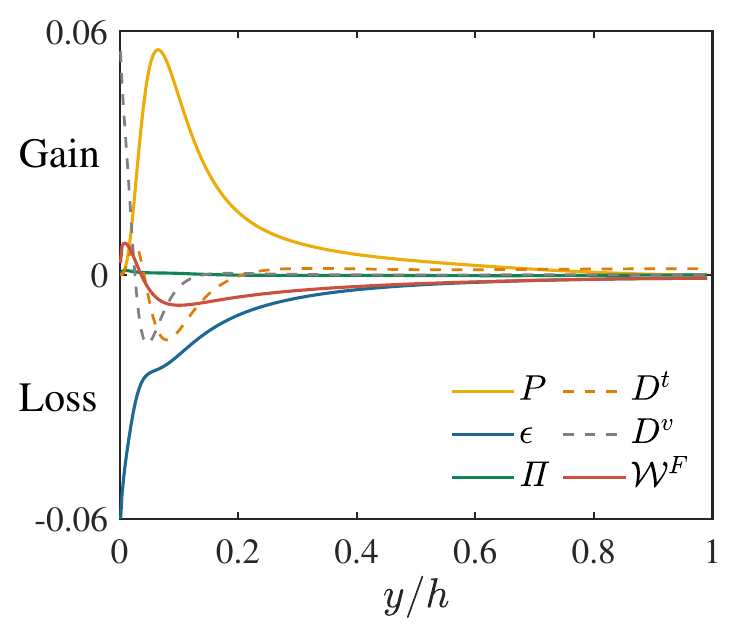}
\put(-1,77){$(b)$}
\put(63,72){$St^+=100$}
\end{overpic}
\begin{overpic}[width=0.48\linewidth]{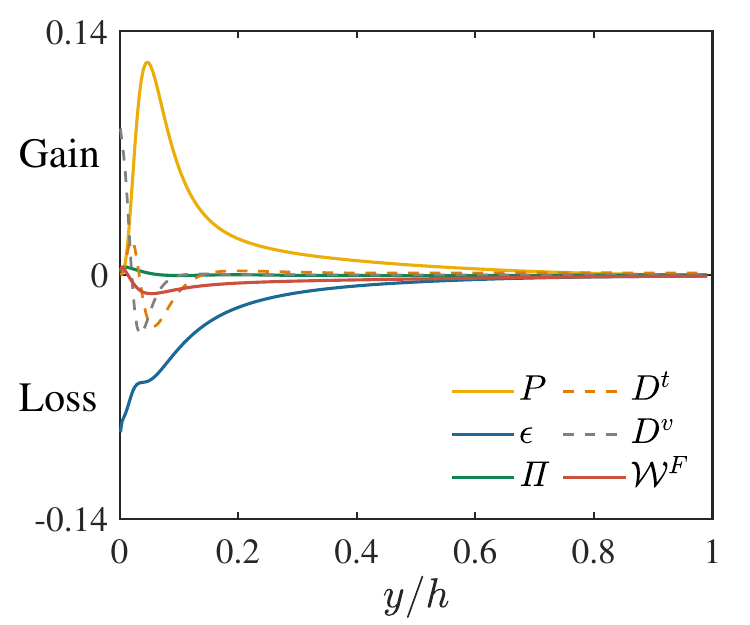}
\put(-1,77){$(c)$}
\put(63,72){$St^+=300$}
\end{overpic}
\begin{overpic}[width=0.48\linewidth]{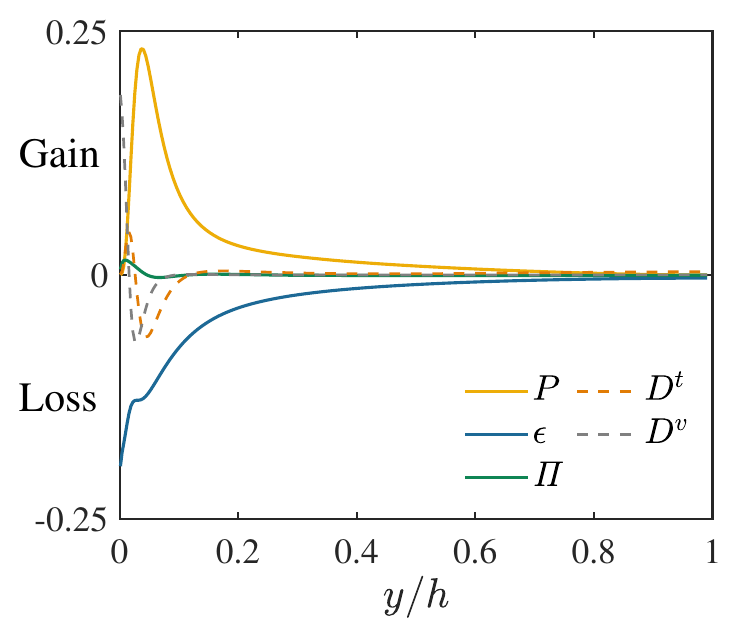}
\put(-1,77){$(d)$}
\end{overpic}
\caption{\label{fig:tke}The TKE budget profiles for three typical cases: $(a)$ St003, $(b)$ St100, and $(c)$ St300, along with $(d)$ the SP results. Hereafter, all the budget terms are scaled by $(u_\tau^{\text{SP}})^4/\nu$.}
\end{figure}
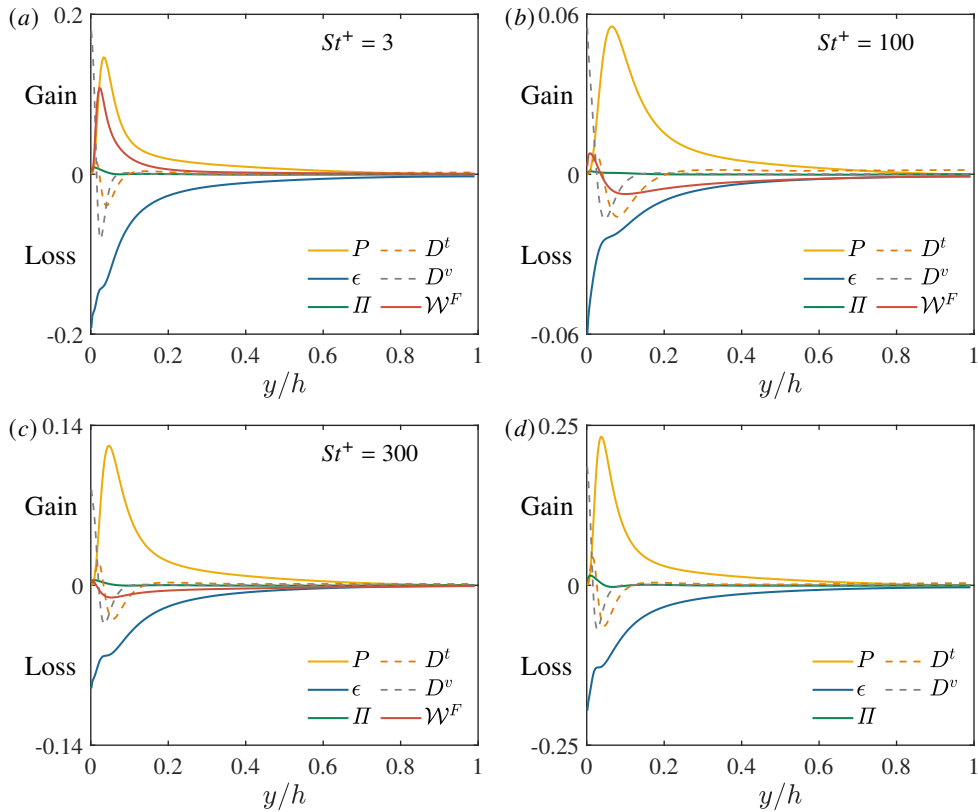

This subsection is directed to examine the turbulent kinetic energy, clarifying how inertial particles affect its production and dissipation.
In fully-developed particle-laden channel flows, the budget equation for TKE ($k=\overline{u'_iu'_i}/2$) reads \citep{Pan2020}
\begin{align}
	0= & \,
	\underbrace{\vphantom{\overline{\frac{\partial u'_i}{\partial x_j}}}-\overline{u'_1u'_2}\frac{\mathrm{d}\bar{u}_1}{\mathrm{d}y}}_{P}
	\underbrace{\vphantom{\overline{\frac{\partial u'_i}{\partial x_j}}}-\nu\overline{\frac{\partial u'_i}{\partial x_j}\frac{\partial u'_i}{\partial x_j}}}_{\epsilon}
	\underbrace{\vphantom{\overline{\frac{\partial u'_i}{\partial x_j}}}-\frac{\mathrm{d}(\overline{u'_iu'_iu'_2}/2)}{\mathrm{d}y}}_{D^t}
	+\underbrace{\vphantom{\overline{\frac{\partial u'_i}{\partial x_j}}}\nu\frac{\mathrm{d}^2 k}{\mathrm{d}y^2}}_{D^v}
	\underbrace{\vphantom{\overline{\frac{\partial u'_i}{\partial x_j}}}-\frac{1}{\rho}\overline{u'_i\frac{\partial p'}{\partial x_i}}}_{\mathit{\Pi}}
	+\underbrace{\vphantom{\overline{\frac{\partial u'_i}{\partial x_j}}}\overline{\mathcal{F}'_iu'_i}}_{{\mathcal{W}}^F}, \label{eqn:tke}
\end{align}
where the right-hand terms are in sequence: $P$ the turbulent production rate; $\epsilon$ the dissipation rate; $D^t$ the turbulent diffusion rate; $D^v$ the viscous diffusion rate; $\mathit{\Pi}$ the velocity pressure-gradient term; and ${\mathcal{W}}^F$ the fluctuating particle-fluid work.
The profiles of these terms are depicted for typical cases in figure~\ref{fig:tke}.
Following previous studies \citep{Lee2015, Gualtieri2023, Cui2024}, special interest is focused on interpreting the key terms, i.e. the turbulent production rate $P$ and the fluctuating particle-fluid work ${\mathcal{W}}^F$.
In comparison with the SP case, it is clear that ${\mathcal{W}}^F$ plays a critical role in the production-dissipation balance.
This term is positive for low-inertia particles and becomes negative in almost the entire channel for high-inertia particles.
Moreover, the `native' turbulent production rate $P$ is reduced for all $St^+$ as compared to the SP case.

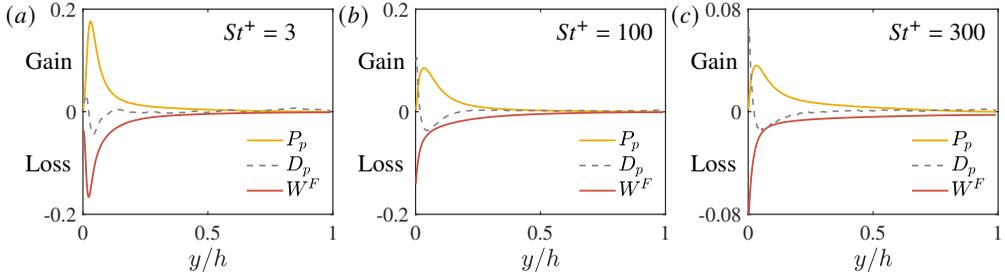
\begin{figure}
\centering
\begin{overpic}[width=0.32\linewidth]{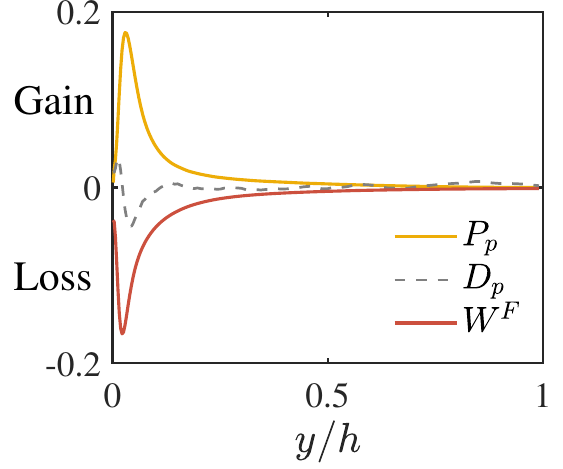}
\put(63,72){$St^+=3$}
\put(-4,77){$(a)$}
\end{overpic}
\begin{overpic}[width=0.32\linewidth]{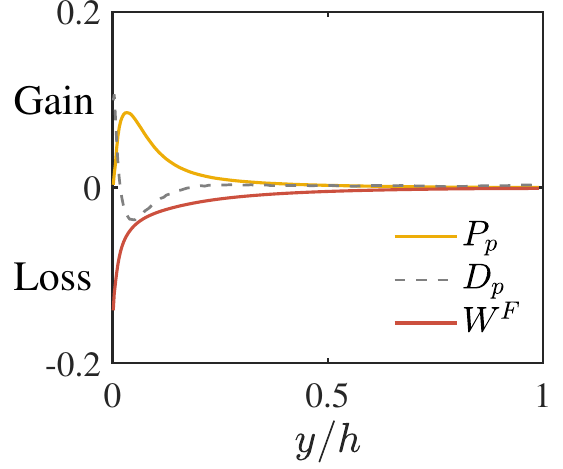}
\put(63,72){$St^+=100$}
\put(-4,77){$(b)$}
\end{overpic}
\begin{overpic}[width=0.32\linewidth]{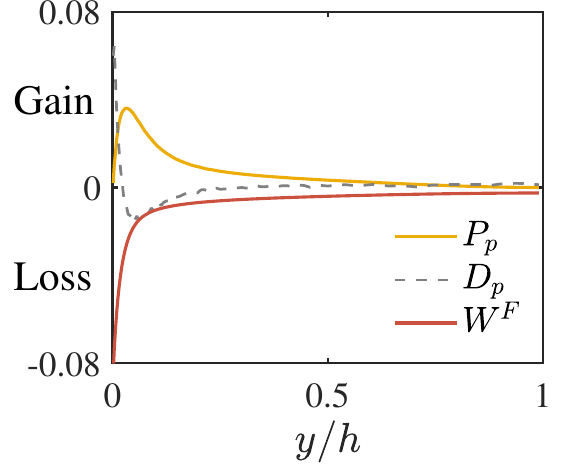}
\put(63,72){$St^+=300$}
\put(-4,77){$(c)$}
\end{overpic}
\caption{\label{fig:pfke}The PFKE budget profiles for three typical cases: $(a)$ St003, $(b)$ St100, and $(c)$ St300.}
\end{figure}

To further understand the role of particle-fluid work on the TKE production/dissipation, we examine the corresponding PFKE ($k_p \equiv \langle v_i^\circ v_i^\circ\rangle/2$) budget (Appendix~\ref{app:stationary}),
\begin{align}
0= & \,
\underbrace{-C\langle v_1^\circ v_2^\circ\rangle \frac{\mathrm{d}\langle v_1\rangle}{\mathrm{d} y}}_{P_p}
\underbrace{-\frac{\mathrm{d}(C\langle v_i^\circ v_i^\circ v_2^\circ \rangle/2)}{\mathrm{d} y}}_{D_p}
+\underbrace{\vphantom{\frac{\mathrm{d}\langle v_1\rangle}{\mathrm{d} y}}C\langle\widehat{F}_i^\circ v_i^\circ\rangle}_{W^F}, \label{eqn:pfke}
\end{align}
where the right-hand terms are identified in sequence as: $P_p$ the particle production rate; $D_p$ the particle diffusion rate; and $W^F$ the fluctuating fluid-particle work.
As demonstrated in figure~\ref{fig:pfke}, the PFKE balance is achieved by the positive-valued $P_p$ which transfers energy from the mean particle motion to the fluctuating counterpart, and the negative-valued work $W^F$ done by carrying fluids.
The particle diffusion $D_p$ serves just to redistribute the PFKE along the wall-normal direction.
Of fundamental importance, the PFKE budget is inherently connected to the TKE budget through the interphase coupling relation (Appendix~\ref{app:coupling}),
\begin{equation}
	{\mathcal{W}}^F=-W^F+N^F=-W^F+\epsilon^{p,F}+\alpha. \label{eqn:nf}
\end{equation} 
Here $N^F$ stands for the net energy loss during fluctuating kinetic energy exchange, $\epsilon^{p,F}=-C\langle\widehat{F}^\circ_i s^\circ_i\rangle$ denotes the fluctuating particle dissipation \citep{Zhao2013}, and $\alpha=C\langle\widehat{F}_i\rangle (\bar{u}_i-\langle u_i\rangle)$ denotes an extra energy transfer term.
Provided the velocity-dependent Stokes drag, the dissipative nature of the $\epsilon^{p,F}$ is straightforward in the physical sense as ${\epsilon}^{p,F}\approx -C\langle s^\circ_is^\circ_i\rangle/\tau\leq0$.
In contrast, $\alpha$ is non-dissipative which transfers kinetic energy between the mean and fluctuating motions involving both fluid and particle phases.
This energy transfer term is nontrivial when there exists a non-zero mean slip velocity accompanied by the preferential accumulation (clustering) of inertial particles \citep{Capecelatro2015,Capecelatro2016}.
According to \citet{Capecelatro2018}, $\alpha$, termed drag production in their paper, becomes dominant only when the mass loading is high enough ${\Phi}_m\geq10$ and some external forcing such as gravity is imposed on particles which leads to a large interphase slip velocity.

\begin{figure}
	\centering
	\begin{overpic}[width=0.48\linewidth]{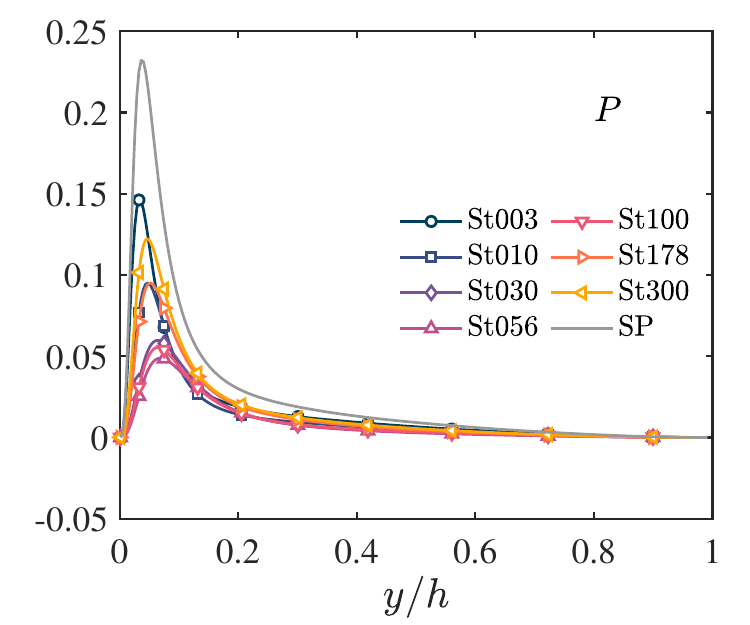}\put(-1,77){$(a)$}\end{overpic}
	\begin{overpic}[width=0.48\linewidth]{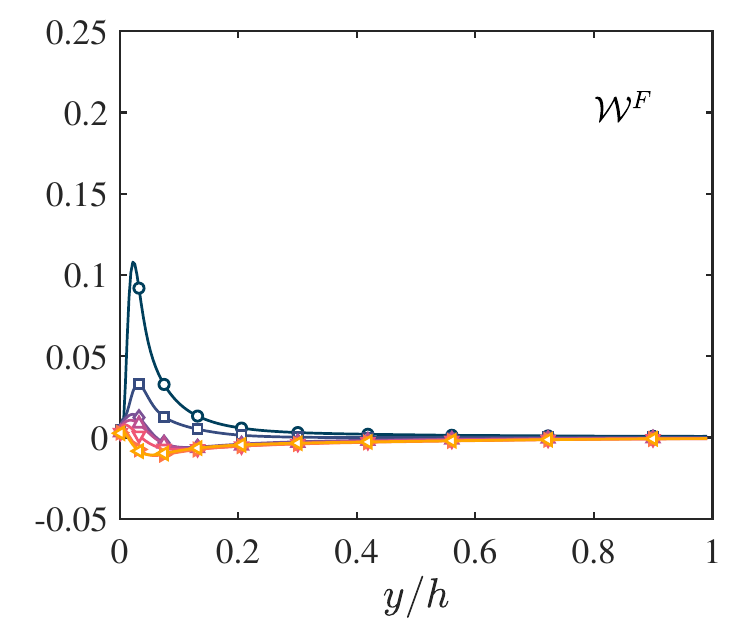}\put(-1,77){$(b)$}\end{overpic}
	\begin{overpic}[width=0.48\linewidth]{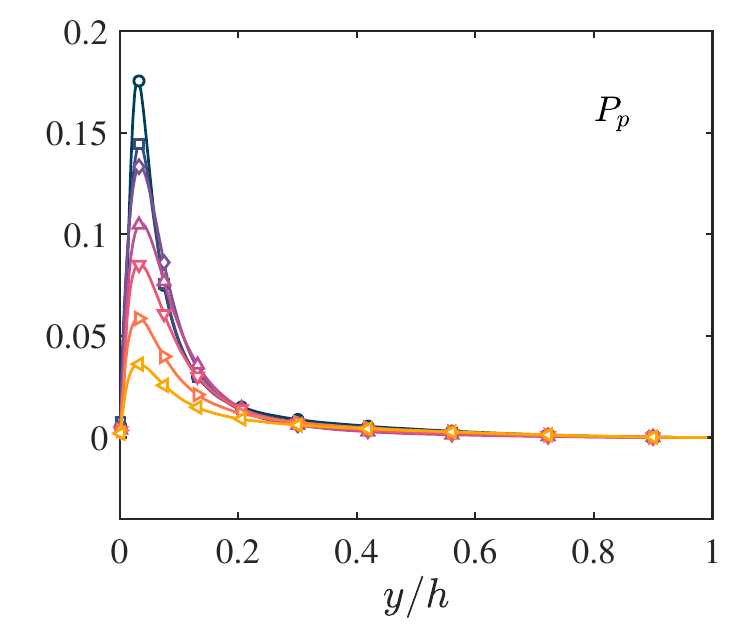}\put(-1,77){$(c)$}\end{overpic}
	\begin{overpic}[width=0.48\linewidth]{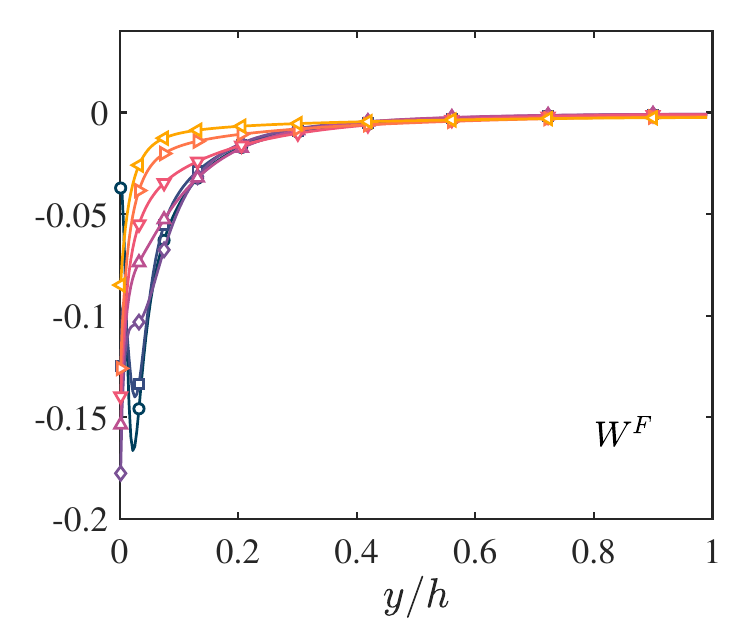}\put(-1,77){$(d)$}\end{overpic}
	\begin{overpic}[width=0.48\linewidth]{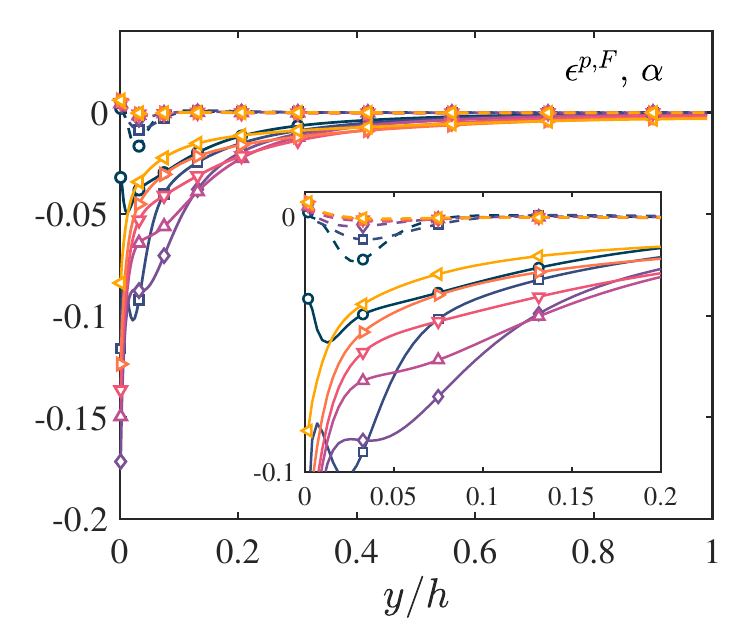}\put(-1,77){$(e)$}\end{overpic}
	\begin{overpic}[width=0.48\linewidth]{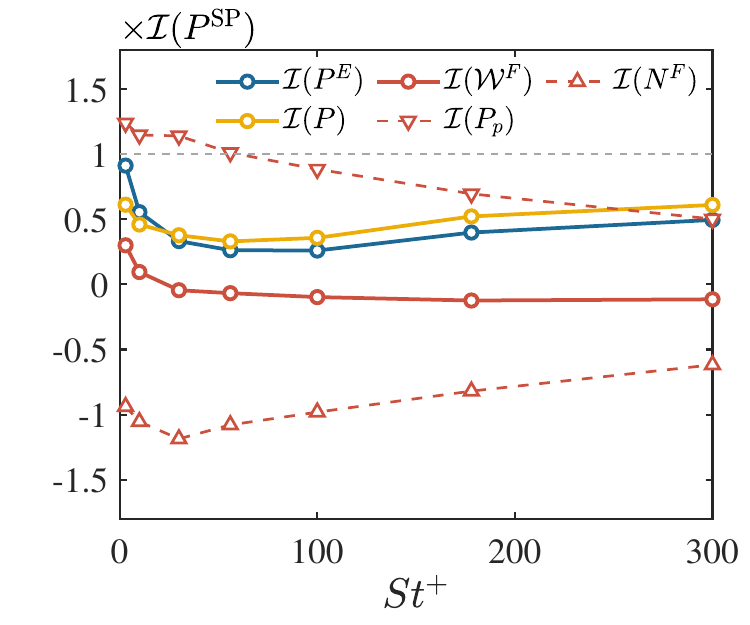}\put(-1,77){$(f)$}\end{overpic}
	\caption{\label{fig:tke-pfke-st}
		The terms that are critical to the TKE, PFKE budgets and TKE-PFKE transfer, including $(a)$ $P$, $(b)$ ${\mathcal{W}}^F$, $(c)$ $P_p$, $(d)$ $W^F$, $(e)$ $\epsilon^{p,F}$ in solid and $\alpha$ in dashed lines, as well as $(f)$ the integrals of dominant terms. 
        In panel~$(e)$, the inset shows a magnified view of the profiles near the wall.
        In panel~$(f)$, $P^E$ denotes the effective production rate which is defined as $P+{\mathcal{W}}^F$, and all the quantities are normalized using the integral of single-phase production rate $\mathcal{I}(P^\text{SP})$.}
\end{figure}

As shown in figure~\ref{fig:tke-pfke-st}, a panoramic view is provided of the dominant terms in Eqs.~(\ref{eqn:tke})-(\ref{eqn:nf}) that contribute to the net energy exchange.
These terms include $P$, ${\mathcal{W}}^F$, $P_p$, $W^F$, $\epsilon^{p,F}$, and $\alpha$.
The Stokes number dependence of each term is clearly illustrated.
The production rate $P$ decreases as $St^+$ increases from $3$ to $56$, before increasing again.
Both $\mathcal{W}^F$ and $P_p$ decreases monotonically, while $W^F$ increases monotonically.
Furthermore, the particle dissipation initially decreases with increasing $St^+$ and then increases, with its extrema obtained at a moderate Stokes number around $30$.
It is noted that compared with ${\epsilon}^{p,F}$, $\alpha$ is indeed negligibly small.
Furthermore, combining Eqs.~(\ref{eqn:pfke}) and (\ref{eqn:nf}) and integrating the terms over the half channel, one obtains the following integral relation,
\begin{equation}
	\mathcal{I}({\mathcal{W}}^F)=\mathcal{I}(P_p)+\mathcal{I}(N^{F})\approx \mathcal{I}(P_p)+\mathcal{I}({\epsilon}^{p,F}). \label{eqn:iwf}
\end{equation}
Here $\mathcal{I}(A)\equiv\int_{0}^{h} A\,\mathrm{d}y$ denotes the wall-normal integral of arbitrary function $A(y)$.
That is to say, the net contribution of the fluctuating particle-fluid work to the TKE budget can be formally decomposed into two dominant components.
As shown in figure~\ref{fig:tke-pfke-st}$(f)$, they include a positive-valued particle production term $\mathcal{I}(P_p)$ which decreases almost monotonically with $St^+$, and a negative-valued particle dissipation term $\mathcal{I}({\epsilon}^{p,F})\approx\mathcal{I}(N^{F})$ which varies non-monotonically with $St^+$.

The underlying mechanisms for the $St^+$ dependence of the above terms are explained as follows.
First, the decreasing of $\mathcal{I}(P_p)$ is attributed to the flattening of the mean particle velocity profile.
As shown in figure~\ref{fig:mean}$(b)$, the mean particle velocity profile flattens as particle inertia increases, leading to a decrease in particle velocity gradient especially in the near-wall region.
Consequently, $\mathcal{I}(P_p)$ decreases monotonically with the continuous increase of $St^+$, considering the fact that $P_p$ is proportional to $\mathrm{d}\langle v_1\rangle/\mathrm{d} y$.
This positive term signifies that inertial particles provide a new way for transferring kinetic energy from the mean flow to turbulent motions through the particle phase, namely, a bypass way termed MKE-PMKE-PFKE-TKE.
Here MKE stands for the mean kinetic energy ($K\equiv \bar{u}_i\bar{u}_i/2$) and PMKE stands for the particle-phase mean kinetic energy ($K_p \equiv \langle v_i\rangle\langle v_i\rangle/2$).
Second, the non-monotonic $St^+$ dependence of particle dissipation $\mathcal{I}({\epsilon}^{p,F})$ can be interpreted by the slip velocity--relaxation time scaling, as indicated by the formulation ${\epsilon}^{p,F}\approx-C\langle s_i^\circ s_i^\circ\rangle/\tau$.
Specifically, when $St^+\to\infty \,(\tau\to\infty)$, the fluctuating slip velocity variance is bound to converge to a finite value, leading to a limit of ${\epsilon}^{p,F}_\infty\to0$.
While $St^+\to0 \,(\tau\to0)$, the leading order of slip velocity is proportional to the relaxation time, $|\pps|\sim O(\tau)$ \citep{Maxey1987}, also leading to ${\epsilon}^{p,F}_0\sim O(\tau^2)/\tau\to0$.
Therefore, it is expected that the fluctuating particle dissipation term exhibits a non-monotonic variation as $St^+$ gradually increases, i.e. initially decreasing from zero, reaching its extrema at a moderate $St^+$, and then recovering back to zero.
Intriguingly, a very recent study offers convincing support to the above analysis \citep{Berk2024}.
Specifically, an analytical model is proposed by \citet{Berk2024} to demonstrate that the slip velocity variance $\langle s^\circ_is^\circ_i\rangle$ follows a piecewise scaling of $\tau^2$--$\tau^1$--$\tau^0$ as $\tau$ increases continuously.
As a consequence of the interplay of these two competing mechanisms, $\mathcal{I}(P_p)+\mathcal{I}({\epsilon}^{p,F})$, the fluctuating particle-fluid work acts as a TKE source for low Stokes numbers while behaves like an energy sink for moderate-to-high Stokes numbers.

\subsection{The self-sustaining processes}
\label{sec:stress}

\begin{figure}
	\centering
	\includegraphics[width=0.72\linewidth]{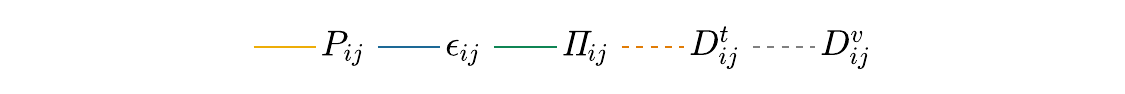}
	\begin{overpic}[width=0.32\linewidth]{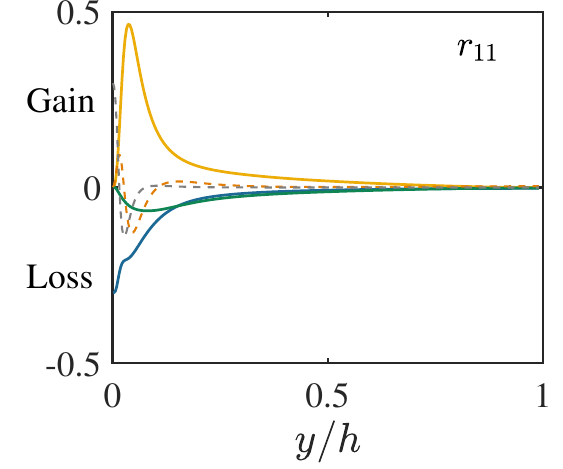}\put(-4,77){$(a)$}\end{overpic}
	\begin{overpic}[width=0.32\linewidth]{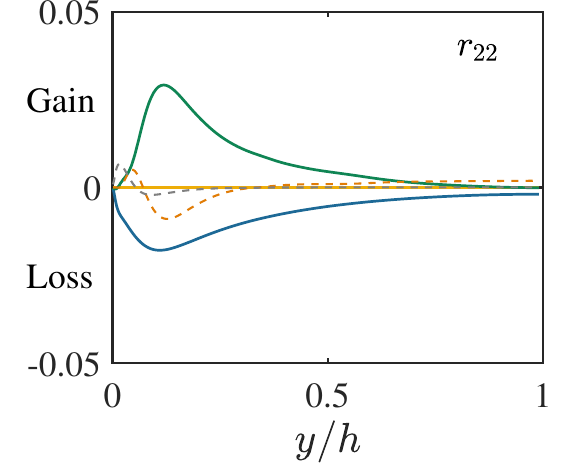}\put(-4,77){$(b)$}\end{overpic}
	\begin{overpic}[width=0.32\linewidth]{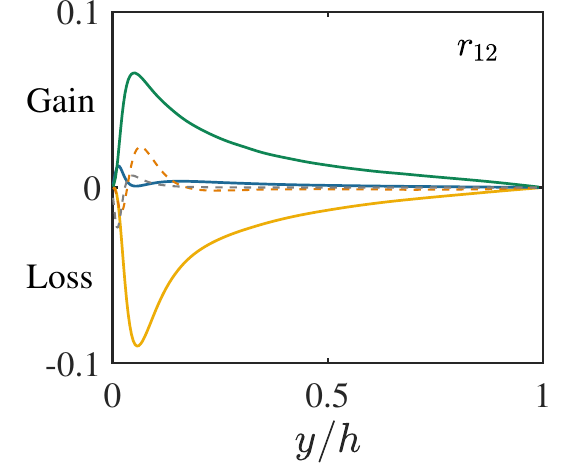}\put(-4,77){$(c)$}\end{overpic}
	\caption{\label{fig:rij-sp}
	The Reynolds stress budgets of the present single-phase channel flow: $(a)$ $r_{11}$, $(b)$ $r_{22}$, $(c)$ $r_{12}$. Similar to the previous section, the budget terms are scaled by $(u_\tau^\text{SP})^4/\nu$.}
\end{figure}

As discussed above, two mechanisms through which particles directly affect the flow properties have been identified by examining the shear stress balance and TKE budget.
The first is the particle-induced extra transport of momentum and kinetic energy, which is positive and manifests as the particle shear stress and the particle production rate.
The second is the particle-induced extra dissipation, which is negative and plays a role in the interphase energy exchange.
Beyond these two, there appears an indirect effect that accounts for the suppression of Reynolds shear stress and turbulent production rate.
To clarify this effect and fulfill the complete physical picture underlying turbulence modulation, we revisit the self-sustaining processes (SSP) of particle-laden channel flows, examining the Reynolds stress ($r_{ij}=\overline{u'_iu'_j}$) budget \citep{Chou1945,Mansour1988},
\begin{align}
	\frac{\partial r_{ij}}{\partial t}= &
	\underbrace{\vphantom{\left(\frac{\partial \bar{u}_i}{\partial x_k}\right)}- \frac{\partial(r_{ij}\bar{u}_k)}{\partial x_k}}_{C_{ij}}
	\underbrace{\vphantom{\left(\frac{\partial \bar{u}_i}{\partial x_k}\right)}-\left(\overline{ u'_j u'_k}\frac{\partial \bar{u}_i}{\partial x_k}+\overline{ u'_i u'_k}\frac{\partial \bar{u}_j}{\partial x_k}\right)}_{P_{ij}}
	\underbrace{\vphantom{\left(\frac{\partial \bar{u}_i}{\partial x_k}\right)}-\frac{\partial\overline{u'_i u'_j u'_k}}{\partial x_k}}_{D^t_{ij}}
	+\underbrace{\vphantom{\left(\frac{\partial \bar{u}_i}{\partial x_k}\right)}\nu\frac{\partial^2r_{ij}}{\partial x_k\partial x_k}}_{D^v_{ij}} \notag \\
	                                                &
	+\underbrace{\frac{1}{\rho}\left(\overline{{u'_i}\frac{\partial p'}{\partial x_j}}+\overline{{u'_j}\frac{\partial p'}{\partial x_i}}\right)}_{\mathit{\Pi}_{ij}}
	+\underbrace{\vphantom{\left(\overline{{u'_j}\frac{\partial p'}{\partial x_i}}\right)}2\nu\overline{\frac{\partial u'_i}{\partial x_k}\frac{\partial u'_j}{\partial x_k}}}_{\epsilon_{ij}} \label{eqn:rsb-sp},
\end{align}
where the right-hand terms are identified in sequence as: $C_{ij}$ the convection term, $P_{ij}$ the production rate, $\epsilon_{ij}$ the dissipation rate, $D^v_{ij}$ the viscous diffusion rate, $D^t_{ij}$ the turbulent diffusion rate, and $\mathit{\Pi}_{ij}$ the velocity pressure-gradient term.
It should be noted that the normal Reynolds stresses are physically equivalent to the components of TKE, given by $k=\overline{u'_iu'_i}/2=r_{ii}/2$.
The SSP in canonical particle-free wall-bounded turbulence has been established as the following three phases \citep{Mansour1988,Pope2000,Zhang2022}: 
1) the Reynolds shear stress $-r_{12}$ acts to pump the kinetic energy of mean motions to $r_{11}$ through $P_{11}$; 
2) part of this kinetic energy is redistributed to $r_{22}$ through $\mathit{\Pi}_{22}$; 
3) in turn, $r_{22}$ works to reproduce $-r_{12}$ through $P_{12}$, closing the TKE regeneration cycle. 
During this cyclic SSP, $r_{33}$ is also generated in the phase of kinetic energy redistribution.
Such an intriguing regeneration cycle of Reynolds stress components in the single-phase channel turbulence is clearly illustrated in figure~\ref{fig:rij-sp}.

\begin{figure}
	\centering
	\includegraphics[width=0.72\linewidth]{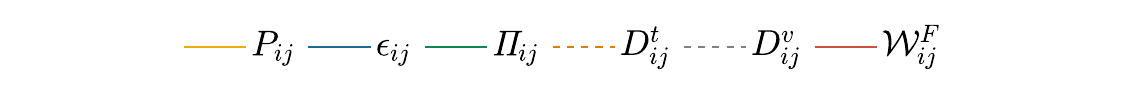}
	\begin{overpic}[width=0.32\linewidth]{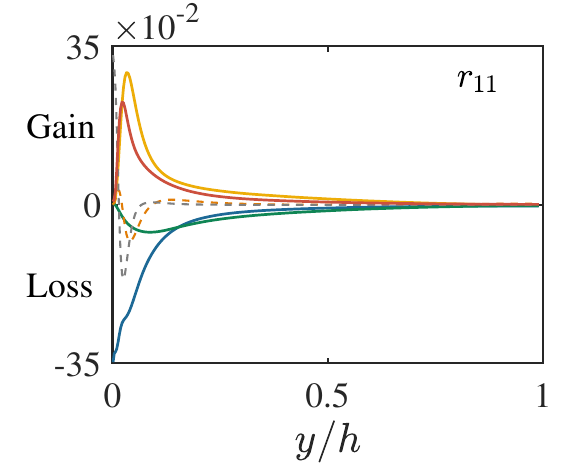}\put(-4,77){$(a)$}\end{overpic}
	\begin{overpic}[width=0.32\linewidth]{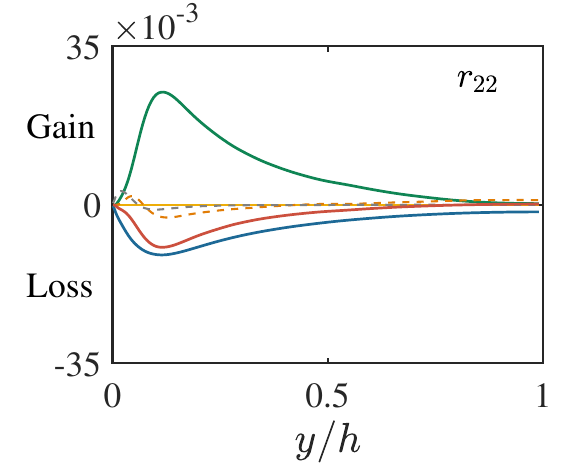}\put(-4,77){$(b)$}\end{overpic}
	\begin{overpic}[width=0.32\linewidth]{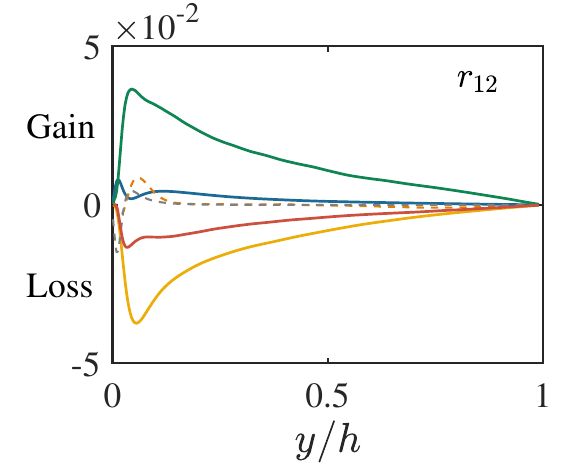}\put(-4,77){$(c)$}\end{overpic}
	\begin{overpic}[width=0.32\linewidth]{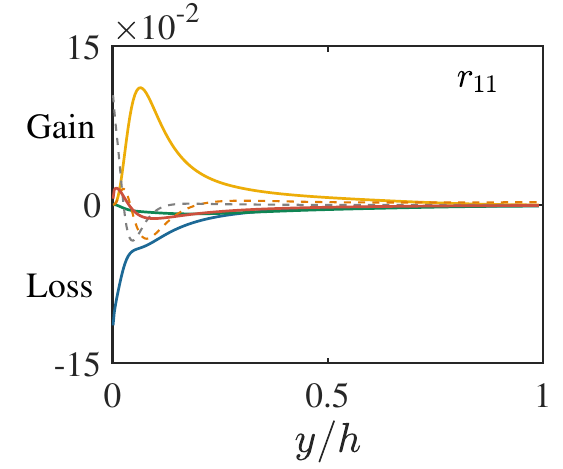}\put(-4,77){$(d)$}\end{overpic}
	\begin{overpic}[width=0.32\linewidth]{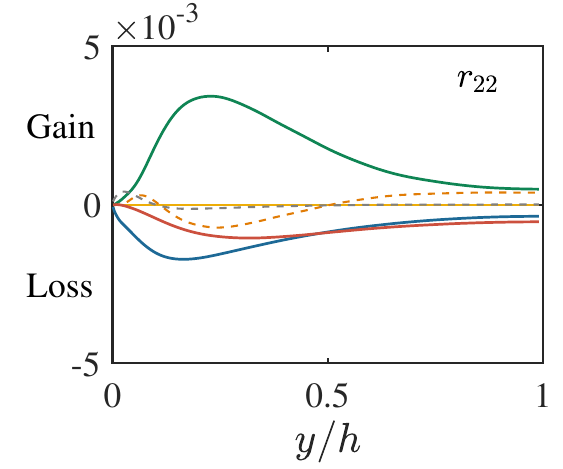}\put(-4,77){$(e)$}\end{overpic}
	\begin{overpic}[width=0.32\linewidth]{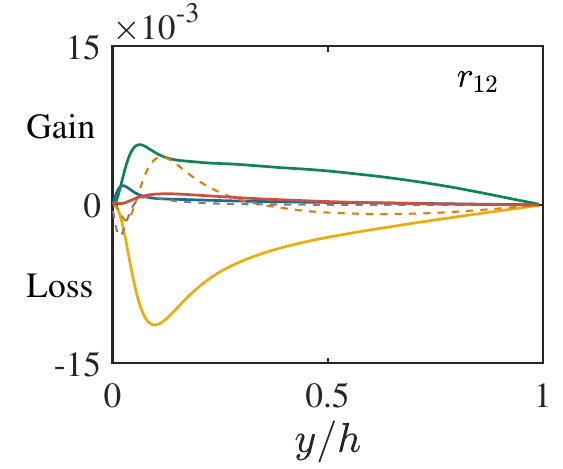}\put(-4,77){$(f)$}\end{overpic}
	\caption{\label{fig:rij}
	The fluid-phase Reynolds stress budgets for $(a, b, c)$ St003 and $(d, e, f)$ St100 cases: $(a, d)$ $r_{11}$, $(b, e)$ $r_{22}$, $(c, f)$ $r_{12}$.}
\end{figure}

For particle-laden channel flow, the Reynolds stress budget can be derived accordingly from the momentum equation (\ref{eqn:flow-momentum}) and reads
\begin{align}
	\frac{\partial r_{ij}}{\partial t}\equiv 0= &
	\underbrace{\vphantom{\left(\frac{\partial \bar{u}_i}{\partial x_k}\right)}-\left(\overline{ u'_j u'_2} \delta_{i1}+\overline{ u'_i u'_2} \delta_{j1}\right) \frac{\partial \bar{u}_1}{\mathrm{d}y} }_{P_{ij}}
	\underbrace{\vphantom{\left(\frac{\partial \bar{u}_i}{\mathrm{d}y}\right)}-\frac{\partial\overline{u'_i u'_j u'_2}}{\mathrm{d}y}}_{D^t_{ij}}
	+\underbrace{\vphantom{\left(\frac{\partial \bar{u}_i}{\mathrm{d}y}\right)}\nu\frac{\partial^2r_{ij}}{\mathrm{d}y^2}}_{D^v_{ij}} \notag \\
	                                                &
	+\underbrace{\frac{1}{\rho}\left(\overline{{u'_i}\frac{\partial p'}{\partial x_j}}+\overline{{u'_j}\frac{\partial p'}{\partial x_i}}\right)}_{\mathit{\Pi}_{ij}}	+\underbrace{\vphantom{\left(\overline{{u'_j}\frac{\partial p'}{\partial x_i}}\right)}2\nu\overline{\frac{\partial u'_i}{\partial x_k}\frac{\partial u'_j}{\partial x_k}}}_{\epsilon_{ij}}    +\underbrace{\vphantom{\left(\overline{{u'_j}\frac{\partial p'}{\partial x_i}}\right)}\overline{\mathcal{F}'_iu_j'}+\overline{\mathcal{F}'_ju_i'}}_{{\mathcal{W}}^F_{ij}},
\end{align}
where an additional term, named fluctuating particle-fluid interaction ${\mathcal{W}}^F_{ij}=\overline{\mathcal{F}'_i u'_j}+\overline{\mathcal{F}'_j u'_i}$, appears on the right-hand side.
In this situation, ${\mathcal{W}}^F_{ij}$ could be viewed as the `input' imposed by particles while other terms could be understood as the `response' of the channel flow system.
From this point, we will mainly focus on the behaviour of ${\mathcal{W}}^F_{ij}$ and the resulting variations in the response.
Each individual term -- the production, diffusion, dissipation, etc. -- will not be addressed one by one because they exhibit a similar response to the particle-fluid interaction as will be shown below.
The budget profiles of the St003 and the St100 cases are depicted in figure~\ref{fig:rij} to demonstrate two typical SSP modulation patterns.
Specifically, for the St003 case, the particle-fluid interaction augments $r_{11}$ and $-r_{12}$ as a source while suppresses $r_{22}$ as a sink.
The extra gains in the streamwise and shear stresses compensate for the loss in the wall-normal stress, allowing the SSP to be maintained with considerable strength.
In contrast, for the St100 case, ${\mathcal{W}}^F_{ij}$ plays a dissipative role in all the Reynolds stress components, resulting in a drastic attenuation of SSP.

\begin{figure}
	\centering
	\includegraphics[width=0.72\linewidth]{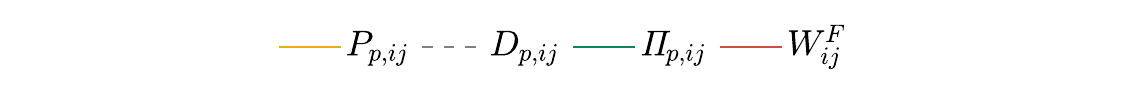}
	\begin{overpic}[width=0.32\linewidth]{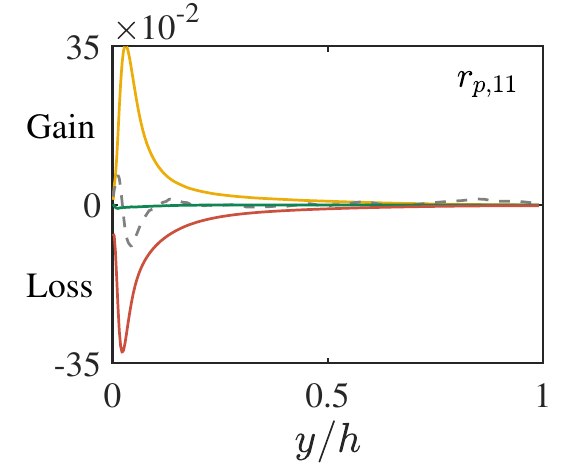}\put(-4,77){$(a)$}\end{overpic}
	\begin{overpic}[width=0.32\linewidth]{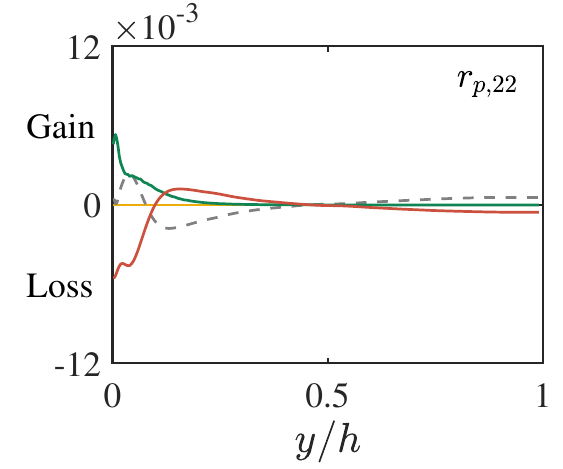}\put(-4,77){$(b)$}\end{overpic}
	\begin{overpic}[width=0.32\linewidth]{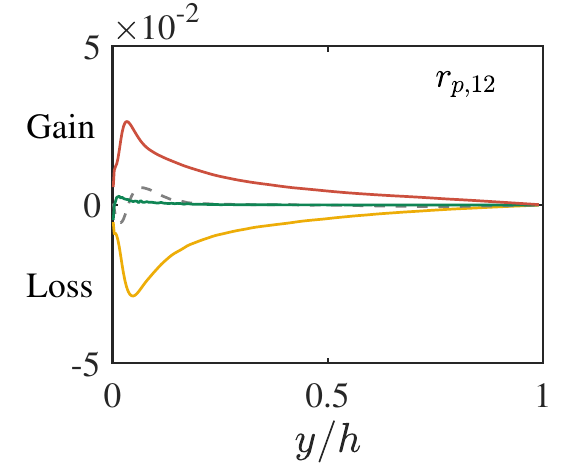}\put(-4,77){$(c)$}\end{overpic}
	\begin{overpic}[width=0.32\linewidth]{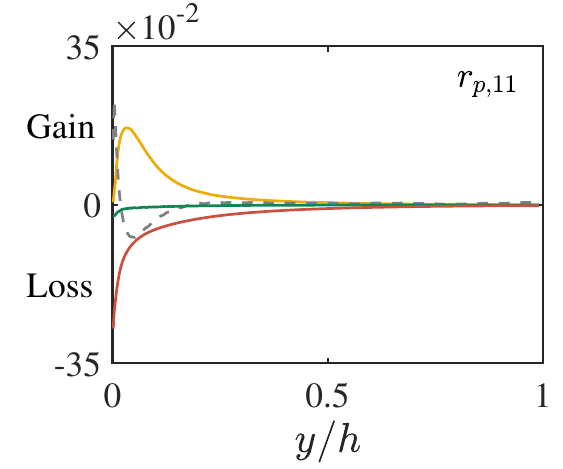}\put(-4,77){$(d)$}\end{overpic}
	\begin{overpic}[width=0.32\linewidth]{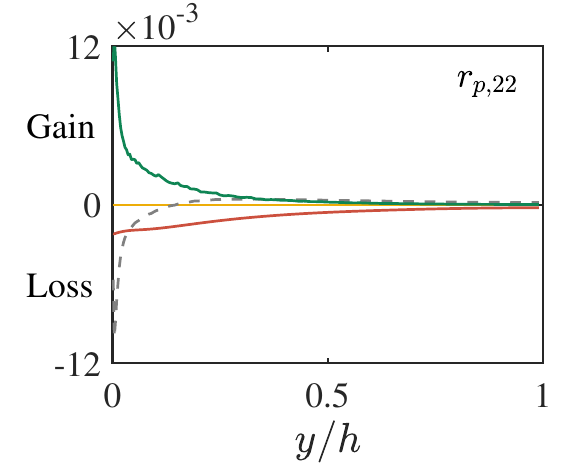}\put(-4,77){$(e)$}\end{overpic}
	\begin{overpic}[width=0.32\linewidth]{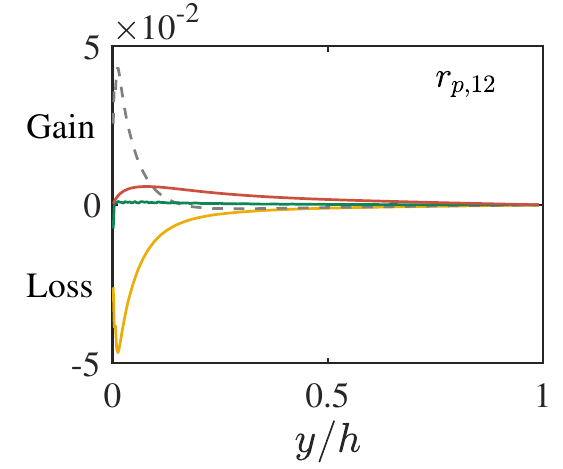}\put(-4,77){$(f)$}\end{overpic}
	\caption{\label{fig:rpij}
	The particle-phase Reynolds stress budgets for $(a, b, c)$ St003 and $(d, e, f)$ St100 cases: $(a, d)$ $r_{p,11}$, $(b, e)$ $r_{p,22}$, $(c, f)$ $r_{p,12}$.}
\end{figure}

Deeper insights into the particle-fluid interactions can be gained by examining the budget equations of particle-phase Reynolds stresses, $r_{p,ij}=\langle v_i^\circ v_j^\circ \rangle$, given as (Appendix~\ref{app:stationary})
\begin{align}
	\frac{\partial(C r_{p,ij})}{\partial t}\equiv0= & \,
	\underbrace{-C \left( \langle v_i^\circ v_2^\circ\rangle \delta_{j1}+ \langle v_j^\circ v_2^\circ\rangle \delta_{i1}\right) \frac{\mathrm{d} \langle v_1\rangle}{\mathrm{d} y} }_{P_{p,ij}}
	\underbrace{\vphantom{\left(\frac{\mathrm{d} \langle v_j\rangle}{\mathrm{d} y}\delta_{j1}\right)}-\frac{\mathrm{d} (C\langle v_i^\circ v_j^\circ v_2^\circ\rangle)}{\mathrm{d} y}}	_{D_{p,ij}}\notag \\
	   & \,+\underbrace{C\left(\langle { \widehat F_i^\circ} v_j^\circ\rangle+\langle { \widehat F_j^\circ} v_i^\circ\rangle\right)}_{W^F_{ij}}
	+\underbrace{\vphantom{C\left(\langle { \widehat F_i^\circ} v_j^\circ\rangle\right)} \{v_i^\circ v_j^\circ\}_\text{coll}}_{\mathit{\Pi}_{p,ij}}, \label{eqn:rpij}
\end{align}
where the right-hand terms are ordered as: $P_{p,ij}$ the particle production rate, $D_{p,ij}$ the particle diffusion rate, $W^F_{ij}$ the fluctuating fluid-particle interaction term, and $\mathit{\Pi}_{p,ij}$ the inter-particle collision term.
Similar to the Reynolds stress, the normal particle-phase Reynolds stresses are physically equivalent to the FPKE components as $k_p \equiv \langle v_i^\circ v_i^\circ\rangle/2=r_{p,ii}/2$.
Corresponding to figure~\ref{fig:rij}, the budget profiles of St003 and St100 cases are provided as typical results in figure~\ref{fig:rpij}.
For $r_{p,11}$, the budget balance is mainly achieved upon a positive particle production rate and a negative fluid-particle interaction term.
A small portion of $r_{p,11}$ is redistributed to the wall-normal and spanwise components $r_{p,22},\,r_{p,33}$ by inter-particle collisions.
For $r_{p,22}$, the particle production disappears due to the absence of mean wall-normal velocity $\langle v_2\rangle$ and its gradient. 
Hence, $\mathit{\Pi}_{p,22}$ is directly balanced by $W^F_{22}$.
For $-r_{p,12}$, the budget is mainly balanced by three elements: particle production, fluid-particle interaction, and particle diffusion.

\begin{figure}
	\centering
	\includegraphics[width=0.72\linewidth]{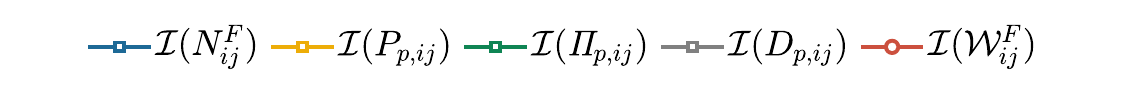}
	\begin{overpic}[width=0.32\linewidth]{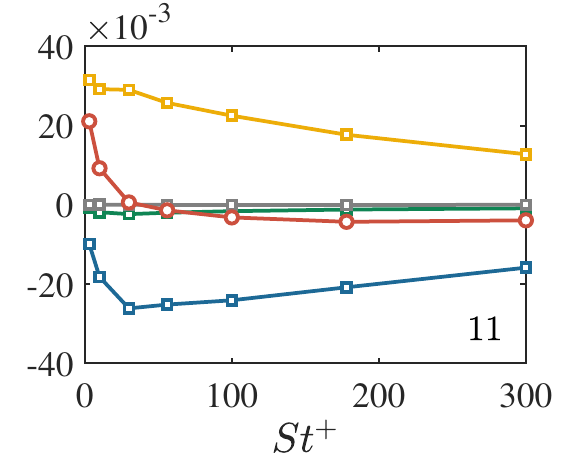}\put(-4,77){$(a)$}\end{overpic}
	\begin{overpic}[width=0.32\linewidth]{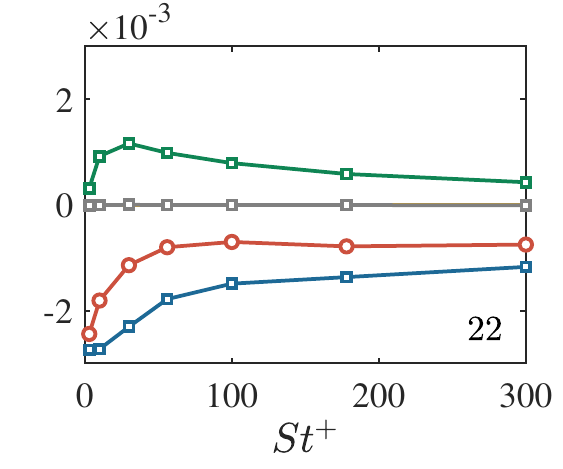}\put(-4,77){$(b)$}\end{overpic}
	\begin{overpic}[width=0.32\linewidth]{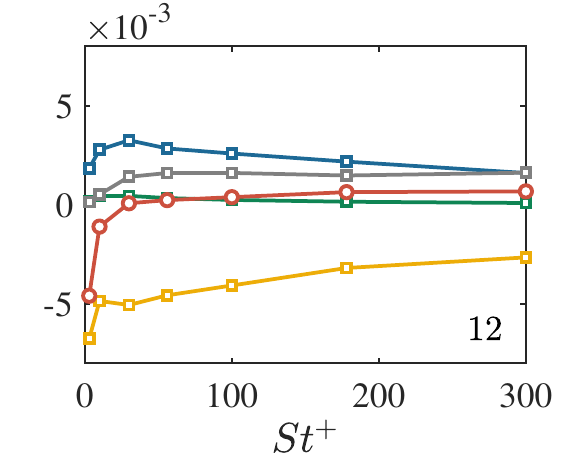}\put(-4,77){$(c)$}\end{overpic}
	\caption{\label{fig:iwfij-st}
	The decomposition of particle-fluid interaction term (\ref{eqn:iwfij}) versus Stokes number: $(a)$ $ij=11$, $(b)$ $ij=22$, $(c)$ $ij=12$.}
\end{figure}

Similar to Eq.~(\ref{eqn:nf}), the interphase coupling relation with regard to each Reynolds stress component can be further characterized as follows
\begin{equation}
	N_{ij}^F=W_{ij}^F+{\mathcal{W}}_{ij}^F =
	{-C\langle\widehat{F}^\circ_i s^\circ_j\rangle-C\langle\widehat{F}^\circ_j s^\circ_i\rangle}
	+{C[\langle\widehat{F}_i\rangle (\bar{u}_j-\langle u_j\rangle)+\langle\widehat{F}_j\rangle (\bar{u}_i-\langle u_i\rangle)]}. \label{eqn:nfij}
\end{equation}
By taking the integral of Eqs.~(\ref{eqn:rpij}) and (\ref{eqn:nfij}), we obtain a decomposition that facilitates the identification of particle-fluid interaction mechanisms, expressed as:
\begin{equation}
	\mathcal{I}({\mathcal{W}}_{ij}^F)=\mathcal{I}(P_{p,ij})+\mathcal{I}(\mathit{\Pi}_{p,ij})+\mathcal{I}(D_{p,ij})+\mathcal{I}(N^{F}_{ij}). \label{eqn:iwfij}
\end{equation}
Figure~\ref{fig:iwfij-st} depicts these integrals as a function of Stokes numbers.
For the $11$ and $12$ components, a pleasing outcome is that the Stokes number dependencies of the two dominant terms, the particle production rate $\mathcal{I}(P_{p,ij})$ and the net energy loss $\mathcal{I}(N^{F}_{ij})$ (with the particle dissipation rate being the dominant contributor), remain consistent with the understanding established in the previous section.
In contrast, the particle diffusion rate plays a secondary role.
Their superposition leads to a shift of ${\mathcal{W}}_{11,12}^F$ from positive to negative as $St^+$ increases.
For the $22$ component, it is shown that the particle dissipation dominates over the collision-induced redistribution throughout the considered parameter space, making ${\mathcal{W}}_{22}^F$ always act as an energy sink.
To sum up, since the particle-induced extra transport requires a mean gradient, ${\mathcal{W}}_{ij}^F$ acts on the carrying flow anisotropically.
The resulting intrinsic anisotropy of particle-laden wall-bounded turbulence has been confirmed in both statistical and structural senses by numerous studies \citep{Zhao2010,Zhao2013,Dave2023,Dritselis2016,Zhou2020,Costa2021}.

\begin{figure}
	\centering
	\includegraphics[width=0.72\linewidth]{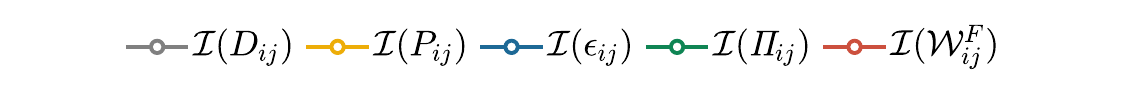}
	\begin{overpic}[width=0.32\linewidth]{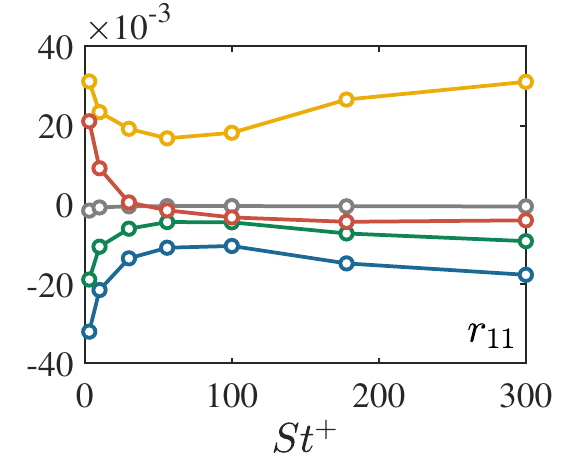}\put(-3,77){$(a)$}\end{overpic}
	\begin{overpic}[width=0.32\linewidth]{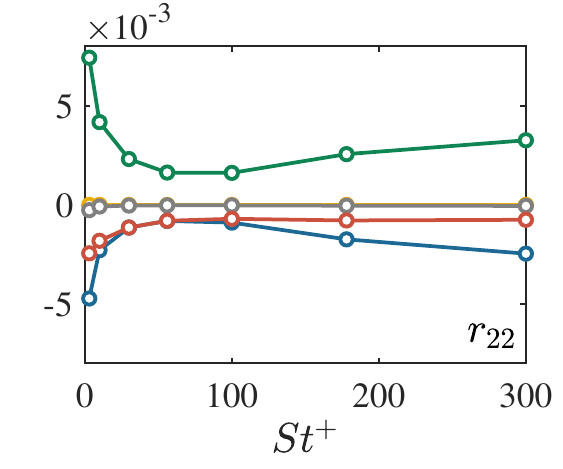}\put(-3,77){$(b)$}\end{overpic}
	\begin{overpic}[width=0.32\linewidth]{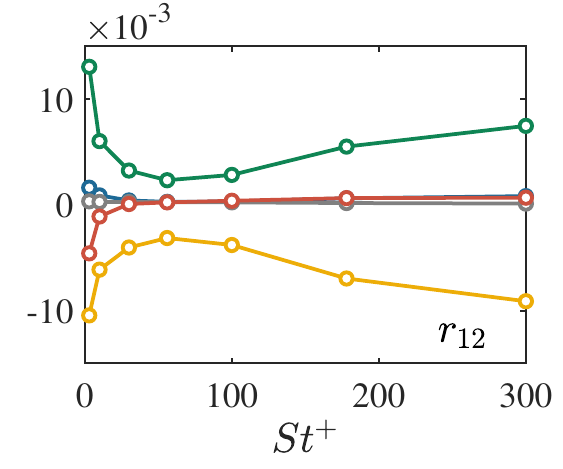}\put(-3,77){$(c)$}\end{overpic}
	\begin{overpic}[width=0.32\linewidth]{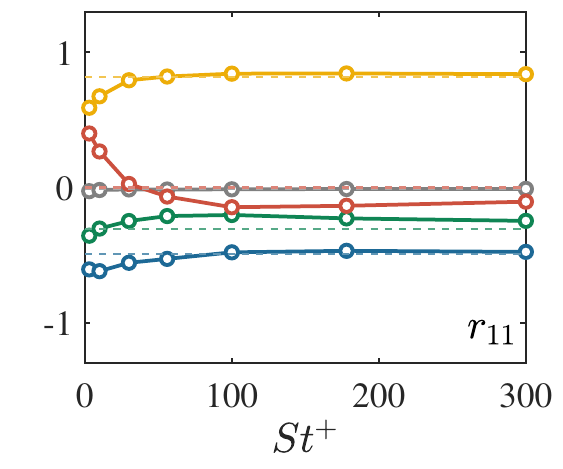}\put(-3,77){$(d)$}\end{overpic}
	\begin{overpic}[width=0.32\linewidth]{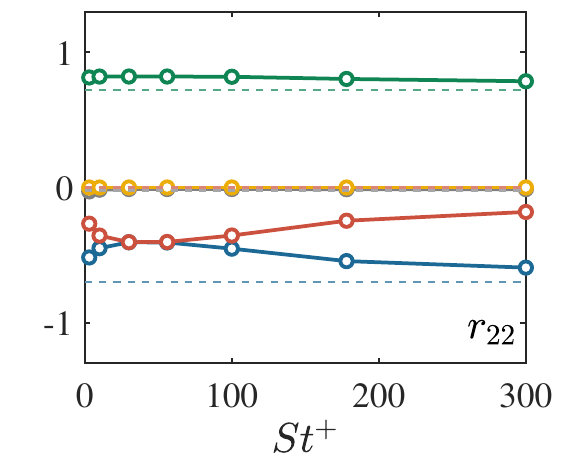}\put(-3,77){$(e)$}\end{overpic}
	\begin{overpic}[width=0.32\linewidth]{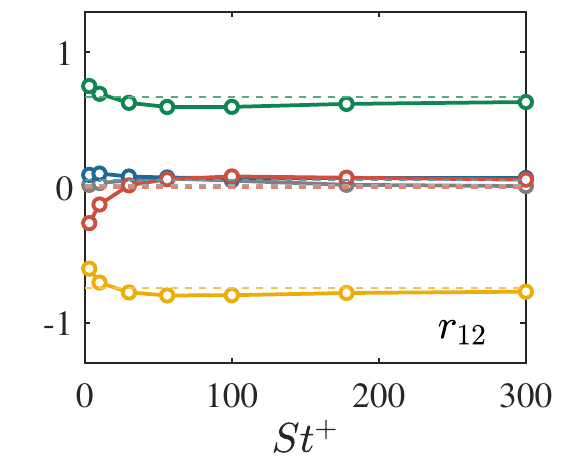}\put(-3,77){$(f)$}\end{overpic}
	\caption{\label{fig:irij-st}
	The Reynolds stress budget integrals: $(a, d)$ $r_{11}$, $(b, e)$ $r_{22}$, $(c, f)$ $r_{12}$. In $(d, e, f)$, the budget integrals are normalized such that, in each case, the summation of their squares equals one.}
\end{figure}

Integrating all the insights discussed above, we turn our attention back to the Reynolds stress budgets and the modification of fluid-phase SSP.
Figure~\ref{fig:irij-st}$(a$-$c)$ depicts the budget integrals versus Stokes numbers.
We observed that almost all the native terms in Reynolds stress budgets respond to the non-monotonic and anisotropic input in a very similar manner.
However, these results contradict the previously established understanding of particle dissipation in the following two aspects.
As shown in figure~\ref{fig:irij-st}$(b)$, the dissipation-dominated particle-fluid interaction in the wall-normal direction, ${\mathcal{W}}^F_{22}$, increases monotonically, rather than initially decreasing before increasing.
Furthermore, at high Stokes numbers, all the particle-fluid interaction components ${\mathcal{W}}^F_{ij}$ remain nearly unchanged, while the amplitudes of the responses, $P_{ij}$, $\epsilon_{ij}$, and $\mathit{\Pi}_{ij}$, increase significantly.

These counterintuitive results are actually attributed to the fluid-particle coupling nature of PLT and will be explained as follows.
As Stokes number (relaxation time) increases from a moderate value, the strength of particle dissipation $|\langle s^\circ_i s^\circ_j\rangle|/\tau$ begins to decrease.
This leads to a weaker suppression of fluid-phase SSP and in turn stronger flow velocity fluctuations.
Notice that $|\langle s^\circ_i s^\circ_j\rangle|/\tau=|\langle (u^\circ_i-v^\circ_i)(u^\circ_j-v^\circ_j)\rangle|/\tau$ depends on the fluctuation of background fluid flow through ${u}_i^\circ$, the absolute strength of particle dissipation should thus be augmented.
In other words, the interphase coupling brings about a `feedback loop' for the flow-velocity-dependent particle dissipation.
Hence, to correctly understand the genuine Stokes number dependence of particle dissipation, one should focus on its relative importance in SSP.
To that end, we redraw the budgets in figure~\ref{fig:irij-st}$(d$-$f)$, where the integrals are normalized such that the summation of their squares equals one \citep[a similar treatment on Reynolds stress budgets can be found in][]{Pope2000}.
Taking the feedback nature into account, we see immediately that the system behaves exactly as expected.
For $r_{11}$ and $-r_{12}$, both extra transport and extra dissipation take effect, leading to the net effect of particle-fluid interaction term varying from positive to negative and finally returning to zero.
For $r_{22}$, the extra dissipation dominates, the negative-valued particle-fluid interaction decreases at first and then increases back to zero.
Moreover, by plotting the normalized integrals in the single-phase case as dashed lines, the system's departure from and return to this reference state in response to the input ${\mathcal{W}}^F_{ij}$ are clearly illustrated.

Overall, the SSP undergoes three distinct stages:
1) at low $St^+$, both $r_{11}$ and $-r_{12}$ are augmented while $r_{22}$ is suppressed (one could expect that, at a lower Stokes number, $St^+<3$, the SSP is likely to be augmented);
2) at moderate $St^+$, the particle dissipation dominates, leading to significant suppression of all Reynolds stress components;
3) at high $St^+$, the suppression effects decline, and the cycling strength recovers, returning to the particle-free state.
This understanding rationalizes the aforementioned indirect effects, as reflected in the non-monotonic modification of the Reynolds shear stress and turbulent production rate.
Up to this point, the particle inertial effects of turbulence modulation at various levels -- momentum transport, kinetic energy transfer, and self-sustaining processes -- have been comprehensively addressed by clarifying both the direct and indirect effects induced by small heavy particles.

\section{Conclusions}
\label{sec:conclusions}

Particle-laden channel flows at a Stokes number ranging widely within $3\leq St^+\leq300$ and a fixed mass loading (${\Phi}_m=0.6$) have been investigated via four-way coupled PP-DNS.
Specifically, a novel flow regime transition (DE-DR-SP) has been substantiated with the continuous increase of $St^+$, highlighting the key role of particle inertia in turbulence modulation.
In the present simulations ($Re_\tau\approx300$), the DE-DR transition occurs at an onset of $St^+\approx10$, with the maximum DR realized at $St^+\approx100$.

For the first time, a comprehensive point-based statistical description of the particle-laden channel turbulence is developed to explore the inter-phase coupling mechanisms.
Taking advantage of this analytical description, the transport processes of this two-phase system are systemically studied, encompassing streamwise momentum transport, kinetic energy transfers, and Reynolds stress budgets.
Through the detailed analysis, two dominant mechanisms that directly modulate the carrying flow dynamics are identified and quantified:
\begin{enumerate}
\item[(1)] a positive mechanism termed particle-induced extra transport, which decreases monotonically with the Stokes number;
\item[(2)] a negative mechanism termed particle-induced extra dissipation, which depends non-monotonically on the Stokes number (decreases at first and then increases, with zero being the two limits).
\end{enumerate}
Their interplay results in either positive or negative contributions of particles on shear stress balance and TKE/Reynolds stress budgets as well as their non-monotonic dependence on the Stokes number.

Of interest, since the extra transport relies on a mean gradient, it does not contribute to the wall-normal Reynolds stress production, thereby introducing an intrinsic anisotropy in inertial particle-rendered turbulence modulation.
Hence, as the Stokes number is increased, the turbulence self-sustaining process undergoes three different stages:
\begin{enumerate}[topsep=2pt,leftmargin=30pt,listparindent=0pt,itemindent=0pt,labelsep=5pt]
\item[(1)] at low $St^+$, particles are found to augment $r_{11}$ and $-r_{12}$ while suppress $r_{22}$;
\item[(2)] at moderate $St^+$, particles act to suppress all the Reynolds stress components;
\item[(3)] at high $St^+$, the particle feedback diminishes, and thus its suppressing effect declines.
\end{enumerate}
These bring together a non-monotonic modulation of SSP strength with regard to the particle inertia: a transition from augmentation to suppression, and ultimately recovering to the particle-free situation.
Consequently, in addition to the direct effects caused by particle-induced extra transport and extra dissipation, an indirect effect arises, which manifests as the non-monotonic modification of Reynolds shear stress and turbulent production rate.
In summary, the competition between modified turbulent transports, extra transports, and extra dissipation, composes a novel understanding of turbulence modulation in particle-laden channel flow, fulfilling a compatible interpretation of the observed flow regime transition (DE-DR-SP) with regard to the Stokes number.
This comprehensive physical picture, which incorporates both fluid- and particle-phase transport processes, is of interest to the flow physics community and could pave the way for future research focused on the mechanistic understanding of particle-laden flows.



\backsection[Declaration of interests]{The authors report no conflict of interest.}

\appendix

\section{Notes on the statistical description}
\label{app:notes}

\subsection{The one-particle probability distribution}
\label{app:notes-distribution}

The one-particle probability distribution introduced in \S~\ref{sec:governing-equations-statistical} can be formally defined as \citep{Johnson2020, Subramaniam2020, Subramaniam2022}
\begin{equation}
	f(\px,\pv;t)\equiv\left\{\delta(\px-\pX(t))\delta(\pv-\pV(t))\right\},
\end{equation}
where $(\pX(t),\pV(t))$ denotes a phase space trajectory of the point particle; $\delta(\,\cdot\,)$ is Dirac delta fucntion (usually approximated by a specific kernel function equipped with the same properties); and $\{\,\cdot\,\}$ means ensemble average over the monodispersed particle system.
Practically, considering a turbulent flow laden with $N_p$ identical particles, we get the following estimation
\begin{align}
f(\px,\pv;t)= & \, {\frac{1}{N_p} \sum_{n=1}^{N_p} \delta(\px-\px_p^{(n)}(t)) \delta(\pv-\pv^{(n)}(t))}, \quad N_p\to\infty. \label{app-eqn:f_estimation}
\end{align}
Here the superscript $(n)$ means the trajectory of the $n$-th particle.
If $N_p$ is large enough, the estimation (\ref{app-eqn:f_estimation}) approaches the true probability distribution, similar to the concept of Kernel density estimation \citep{Epanechnikov1969}.

\subsection{Derivation of the transport equations}
\label{app:derivation}

The phase-space transport equation (\ref{eqn:Boltzmann}), when multiplied by a function $A(\px,\pv)$ and applied with the operator $\Phi_m L_y \idotsint \cdot \,\mathrm{d}\pv \mathrm{d}x \mathrm{d}z$, yields
\begin{equation}
\frac{\partial (C \langle A \rangle)}{\partial t} + \frac{\pat  (C\langle Av_2\rangle)}{\pat y}-C\left\langle v_k\frac{\partial A}{\partial x_k}\right\rangle
-C\left\langle \widehat{F}_k\frac{\partial A}{\partial v_k}\right\rangle
= \{ A \}_\text{coll}, \label{app-eqn:mean-transport}
\end{equation}
where $\widehat{F}_k=F_k/m_p$,
\beq\lb{eq.C1}
C(y,t) = \Phi_m L_y \idotsint f(\px,\pv;t)\,\mathrm{d} \pv \mathrm{d}x \mathrm{d}z,
\eeq 
and 
\begin{equation}\label{Acoll}
\{ A \}_\text{coll} = \Phi_m L_y \idotsint \, A \left( \frac{\partial f}{\partial t} \right)_\text{coll} \mathrm{d}\pv\mathrm{d}x\mathrm{d}z . 
\end{equation}
Substituting $A=1,\,v_i,\,v_iv_i/2$ into Eq.~(\ref{app-eqn:mean-transport}) respectively, regarding that an arbitrary constant, momentum, and kinetic energy are collisional invariants \citep{Vincenti1965},  
we obtain
\beqn
&& \fr{\pat C }{\pat t}+ \fr{ \pat (C\langle v_2\rangle)}{ \pat y} =0 , \label{app-eqn:pc}   \\
&& \frac{\partial (C \langle v_i \rangle)}{\partial t} + \frac{\pat (C\langle v_i v_2\rangle)}{\pat y} - C\langle\widehat{F}_i\rangle =0,  \label{app-eqn:pm}  \\
&& \frac{\partial (C \langle v_i v_i \rangle /2)}{\partial t}+ \frac{\pat (C\langle v_iv_i v_2\rangle/2)}{\pat y} -C\langle \widehat{F}_i v_i \rangle = 0. \label{app-eqn:pke}
\eeqn
Clearly, Eqs.\,(\ref{app-eqn:pc})-(\ref{app-eqn:pke}) represent the conservation of mass, momentum, and kinetic energy of the particle phase per unit mass.
Note that, Eq.~\er{app-eqn:pc} and the $y$-component of Eq.~\er{app-eqn:pm} are exactly the same as Eqs.~(3.5) and (3.7) presented in \citet{Johnson2020}.

From Eqs.~\er{app-eqn:pc} and \er{app-eqn:pm}, the equation of particle-phase mean kinetic energy (PMKE, $K_p= \langle v_i\rangle\langle v_i\rangle/2$) reads
\beq\label{app-eqn:pmke}
\frac{\partial (C K_p)}{\partial t}  
    =- \frac{\partial (C K_p \langle v_2 \rangle)}{\partial y} 
    + C\langle v_i^\circ v_2^\circ\rangle\frac{\partial \langle v_i \rangle}{\partial y} 
    - \frac{\partial (C\langle v_i^\circ v_2^\circ\rangle\langle v_i \rangle)}{\partial y} 
    + C\langle\widehat{F}_i\rangle\langle v_i \rangle.
\eeq
Here $\langle v_i v_i\rangle = \langle v_i \rangle  \langle v_i \rangle +  \langle v_i^\circ  v_i^\circ \rangle$ and $\langle \widehat{F}_i v_i\rangle = \langle\widehat{F}_i\rangle \langle v_i \rangle + \langle \widehat{F}_i^\circ v_i^\circ \rangle$. Taking together Eqs.~\er{app-eqn:pke} and \er{app-eqn:pmke} yields the equation of particle-phase fluctuating kinetic energy (PFKE, $k_p = \langle v_i^\circ v_i^\circ \rangle/2$), reading as
\beq\label{app-eqn:pfke}
\frac{\partial (C k_p)}{\partial t} =
    - \frac{\partial (C k_p \langle v_2 \rangle)}{\partial y} 
    - C\langle v_i^\circ v_2^\circ\rangle\frac{\partial \langle v_i \rangle}{\partial y} 
    - \frac{\partial (C\langle v_i^\circ v_i^\circ v_2^\circ\rangle/2)}{\partial y} 
    + C\langle \widehat{F}^\circ_i v^\circ_i\rangle.
\eeq

Substituting $A=v_iv_j$ into Eq.~(\ref{app-eqn:mean-transport}), we have
\beq\lb{app-eqn:vivj}
\fr{\pat (C \langle v_i v_j \rangle)}{\pat t} = -\fr{\pat (C\langle v_i v_j v_2\rangle)}{\pat y} + C (\langle \widehat{F}_i v_j \rangle + \langle \widehat{F}_j v_i \rangle) + \{ v_i v_j \}_\text{coll}.
\eeq
Here, since the inter-particle collision leads to a redistribution of kinetic energy components among the three directions, the collision term $\{\,\cdot\,\}_\text{coll}$ does not vanish.
In practice, when solving the inter-particle collision using hard-sphere model, one can obtain the difference in second-order moments of the pre-collision and post-collision velocities at each time step to compute $\{v_iv_j\}_\text{coll}$.
Again, using Eqs.~\er{app-eqn:pc} and \er{app-eqn:pm}, the transport equation of the second-order moments of mean particle velocities is obtained as
\beq\lb{eq.vivj-mean}
\begin{split}
\fr{\pat (C \langle v_i \rangle \langle v_j \rangle)}{\pat t} = & - \fr{ \pat ( C \langle v_i \rangle \langle v_j \rangle \langle v_2\rangle)}{ \pat y} + C \left( \langle v_i^\circ v_2^\circ \rangle \frac{\pat \langle v_j \rangle}{\pat y} + \langle v_j^\circ v_2^\circ \rangle \frac{\pat \langle v_i \rangle}{\pat y} \right) \\
& - \frac{\pat [C( \langle v_i^\circ v_2^\circ \rangle \langle v_j \rangle + \langle v_j^\circ v_2^\circ \rangle \langle v_i \rangle)]}{\pat y} + C (\langle \widehat{F}_i \rangle \langle v_j \rangle + \langle \widehat{F}_j \rangle \langle v_i \rangle).
\end{split}
\eeq
Clearly, by setting $v_i = v_j$ and applying the Einstein summation convention, Eq.~\er{eq.vivj-mean} reproduces Eq.~\er{app-eqn:pmke}.
Subtracting Eq.~\er{eq.vivj-mean} from Eq.~(\ref{app-eqn:vivj}) yields the transport equation of the second-order moments of fluctuating velocities, given as
\beq\lb{eq.vivj-fluc}
\begin{split}
\fr{\pat (C \langle v_i^\circ v_j^\circ \rangle)}{\pat t} =& - \fr{ \pat (C \langle v_i^\circ v_j^\circ \rangle \langle v_2\rangle)}{ \pat y} - C \left( \langle v_i^\circ v_2^\circ \rangle \frac{\pat \langle v_j \rangle}{\pat y} + \langle v_j^\circ v_2^\circ \rangle \frac{\pat \langle v_i \rangle}{\pat y} \right) \\
& - \frac{\pat ( C \langle v_i^\circ v_j^\circ v_2^\circ \rangle)}{\pat y} + C (\langle \widehat{F}_i^\circ v_j^\circ \rangle + \langle \widehat{F}_j^\circ v_i^\circ \rangle) + \{ v^\circ_i v^\circ_j \}_\text{coll},
\end{split}
\eeq
where $\{ v^\circ_i v^\circ_j \}_\text{coll}= \{  v_i v_j \}_\text{coll}$ according to Eq.~(\ref{Acoll}).

\subsection{Statistically stationary state}
\label{app:stationary}

After the whole system reaches the statistically stationary states for both the fluid and the particle phases, $\pat \langle \, \cdot \, \rangle / \pat t = \pat \overline{( \, \cdot \, )} / \pat t =0$, Eqs.~\er{app-eqn:pc}-\er{app-eqn:pke} reduce to 
\beqn
&& \fr{ {\rm d} (C\langle v_2\rangle)}{{\rm d} y} =0 , \label{app-eqn:pc-stationary}   \\
&& \frac{{\rm d} (C\langle v_i v_2\rangle)}{{\rm d} y} - C\langle\widehat{F}_i\rangle =0,  \label{app-eqn:pm-stationary}  \\
&& \frac{{\rm d} (C\langle v_iv_i v_2\rangle/2)}{{\rm d} y} -C\langle \widehat{F}_i v_i \rangle = 0. \label{app-eqn:pke-stationary}
\eeqn
Given the concentration is positive semi-definite, the particle continuity equation leads to a vanished mean wall-normal drift $\langle v_2\rangle\equiv0$. 
The periodicity and symmetry of the system lead to $\langle v_3\rangle\equiv0$, so the mean particle velocity reduces to $\langle \pv\rangle=(\langle v_1\rangle(y),\,0,\,0)$. In that, the equations of PMKE and PFKE reduce to
\beq\label{app-eqn:pmke-stationary}
0=C\langle v_1^\circ v_2^\circ \rangle \frac{{\rm d} \langle v_1 \rangle}{{\rm d} y} - \frac{{\rm d} (C\langle v_1^\circ v_2^\circ \rangle  \langle v_1 \rangle)}{{\rm d} y} + C\langle\widehat{F}_1 \rangle \langle v_1 \rangle, 
\eeq
and
\beq \label{app-eqn:pfke-stationary}
0=\underbrace{-C\langle v_1^\circ v_2^\circ\rangle \frac{\mathrm{d}\langle v_1\rangle}{\mathrm{d} y}}_{P_p}
\underbrace{-\frac{\mathrm{d}(C\langle v_i^\circ v_i^\circ v_2^\circ \rangle/2)}{\mathrm{d} y}}_{D_p}
+\underbrace{\vphantom{\frac{\mathrm{d}\langle v_1\rangle}{\mathrm{d} y}}C\langle\widehat{F}_i^\circ v_i^\circ\rangle}_{W^F}. 
\eeq
The terms in the PFKE budget are the particle production rate $P_p$, the particle diffusion rate $D_p$, and the fluctuating fluid-particle work $W^F$, respectively. 

Accordingly, the transport equation for second-order moments of mean and fluctuating particle velocities reduce to
\begin{align}
0=& \, C ( \langle v_i^\circ v_2^\circ\rangle \delta_{j1} + \langle v_j^\circ v_2^\circ \rangle \delta_{i1} ) \frac{\mathrm{d} \langle v_1 \rangle}{\mathrm{d} y}
-\frac{ {\rm d} [ C ( \langle v_i^\circ v_2^\circ\rangle \delta_{j1} + \langle v_j^\circ v_2^\circ\rangle \delta_{i1} ) \langle v_1 \rangle ] }{{\rm d} y} \notag \\
& \,+ C ( \langle\widehat{F}_i\rangle \delta_{j1} + \langle\widehat{F}_j\rangle \delta_{i1} ) \langle v_1 \rangle. \lb{eq.vivj-fluc-stationary}
\end{align}
and
\begin{align}
0=& \,\underbrace{ - C ( \langle v_i^\circ v_2^\circ\rangle \delta_{j1}+ \langle v_j^\circ v_2^\circ\rangle \delta_{i1} ) \frac{\mathrm{d} \langle v_1 \rangle}{\mathrm{d} y} }_{P_{p,ij}}	
\underbrace{\vphantom{\left(\frac{\mathrm{d} \langle v_j\rangle}{\mathrm{d} y}\delta_{j1}\right)} 
-\frac{\mathrm{d} (C\langle v_i^\circ v_j^\circ v_2^\circ\rangle)}{\mathrm{d} y}}_{D_{p,ij}}\notag \\
& \,+\underbrace{\vphantom{}C (\langle { \widehat F_i^\circ} v_j^\circ\rangle+\langle { \widehat F_j^\circ} v_i^\circ\rangle )}_{W^F_{ij}} 
+\underbrace{\vphantom{C\left(\langle { \widehat F_i^\circ} v_j^\circ\rangle \right)} \{v_i^\circ v_j^\circ\}_\text{coll}}_{\mathit{\Pi}_{p,ij}}. \lb{eq.vivj-fluc-stationary}
\end{align}
Regarding the similarity between $\langle v_i^\circ v_j^\circ\rangle$ and $\overline{u'_iu'_j}$, we term it the particle-phase Reynolds stress. The meaning of each term in the particle-phase Reynolds stress budget is identified as: $P_{p,ij}$ the particle production rate, $D_{p,ij}$ the particle diffusion rate, $W^F_{ij}$ the fluctuating fluid-particle interaction term, and $\mathit{\Pi}_{p,ij}$ the inter-particle collision term.

\subsection{The interphase coupling relations}\label{app:coupling}

Since the momentum and energy balance of the particle-laden channel flow is achieved by the tight coupling between the two phases, all the above transport equations can be directly connected with their fluid-phase counterparts.
Starting from Eqs.~(\ref{app-eqn:f_estimation}) and (\ref{eqn:feedback}), we have
\begin{align}
C \langle \widehat{F}_i \rangle 
= & \, \Phi_m L_y \idotsint \, \widehat{F}_i f \mathrm{d}\pv \mathrm{d}x \mathrm{d}z \\
= & \, \frac{\Phi_m L_y}{N_p m_p} \int_{0}^{L_z} \!\!\!\! \int_{0}^{L_x}\sum_{n=1}^{N_p} \delta(\px-\px_p^{(n)}(t)) F_i^{(n)} \mathrm{d}x \mathrm{d}z \\
= & \, \frac{1}{\rho L_xL_z} \int_{0}^{L_z} \!\!\!\! \int_{0}^{L_x}\sum_{n=1}^{N_p} \delta(\px-\px_p^{(n)}(t)) F_i^{(n)} \mathrm{d}x \mathrm{d}z \\
= & \, - \overline{\mathcal{F}}_i,
\end{align}
where $\Phi_m= (N_p m_p) / (\rho L_x L_y L_z)$ and the sampling property of the Dirac delta function is utilized.
Notice that, the Reynolds (namely, the ensemble) average in the statistically one-dimensional channel flow is estimated by the streamwise and spanwise averages, allowing us to arrive at the final step.
By a similar way, we have
\begin{equation}
{{\mathcal{W}}} = \overline{\mathcal{F}_i {u}_i} = -C\langle\widehat{F}_iu_i\rangle.    \label{app-eqn:coupling-w}
\end{equation}

For the sake of brevity, the following notations are adopted for result discussion
\begin{align}
W\equiv  & \, C\langle \widehat{F}_i v_i\rangle,  & {\mathcal{W}}\equiv & \, \overline{\mathcal{F}_iu_i}; \label{eqn:fp-pf} \\
W^M\equiv & \, C\langle \widehat{F}_i\rangle\langle v_i\rangle, & {\mathcal{W}}^M\equiv & \, \overline{\mathcal{F}}_i{\bar{u}_i}; \label{eqn:fpm-pfm} \\
W^F\equiv & \, C\langle \widehat{F}^\circ_iv^\circ_i\rangle,    & {\mathcal{W}}^F\equiv & \, \overline{\mathcal{F}'_iu'_i},\label{eqn:fpf-pff}
\end{align}
where the superscript $M$ and $F$ stand for mean-motion and fluctuating components, respectively; and identities $W=W^M+W^F$ and ${\mathcal{W}}={\mathcal{W}}^M+{\mathcal{W}}^F$ hold since  $\langle v_i^\circ\rangle=0$ and $\overline{u_i'}=0$.
The net loss of kinetic energy during interphase coupling is then derived as follows:
\begin{align}
N=    & \,W+{\mathcal{W}}= {-C\langle\widehat{F}_i s_i\rangle}, \label{app-eqn:n} \\
N^M = & \, W^M+{\mathcal{W}}^M =
{-C\langle\widehat{F}_i\rangle \langle s_i\rangle}
-{C\langle\widehat{F}_i\rangle (\bar{u}_i-\langle u_i\rangle)}, \label{app-eqn:nm}       \\
N^F = & \, W^F+{\mathcal{W}}^F =
\underbrace{ -C\langle\widehat{F}^\circ_i s^\circ_i\rangle }_{ \ep^{p,F} } +
\underbrace{\vphantom{C\langle\widehat{F}^\circ_i s^\circ_i\rangle} {C\langle\widehat{F}_i\rangle (\bar{u}_i-\langle u_i\rangle)} }_{ \al }, \label{app-eqn:nf}
\end{align}
where $s_i=u_i-v_i$ is the interphase slip velocity, $\epsilon^{p,F}$ denotes the fluctuating particle dissipation \citep{Zhao2013}, and $\alpha$ denotes an extra energy transfer term.

\section{Validation}
\label{app:validation}

\begin{figure}
    \centering
    \begin{overpic}[width=0.32\linewidth]{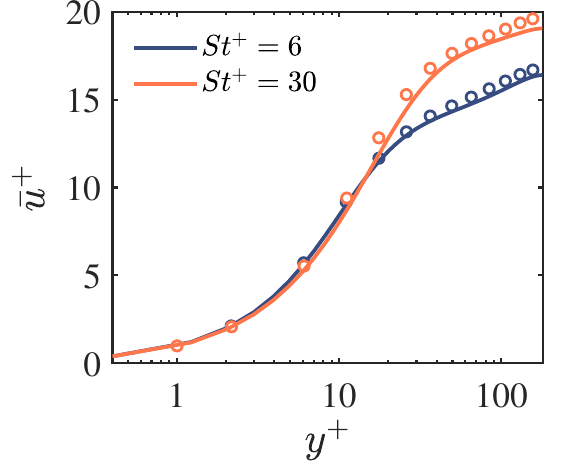}\put(-3,77){$(a)$}\end{overpic}
    \begin{overpic}[width=0.32\linewidth]{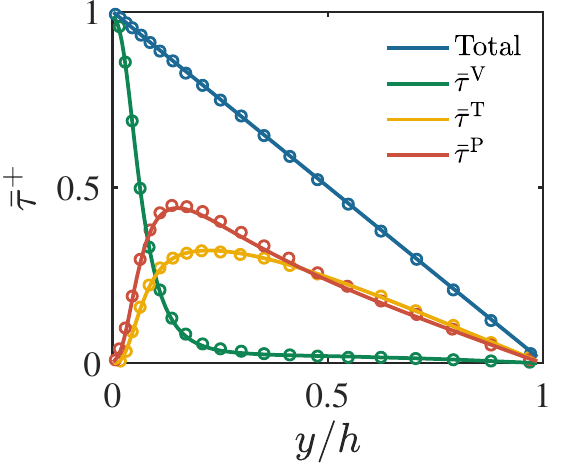}\put(-3,77){$(b)$}\end{overpic}
    \begin{overpic}[width=0.32\linewidth]{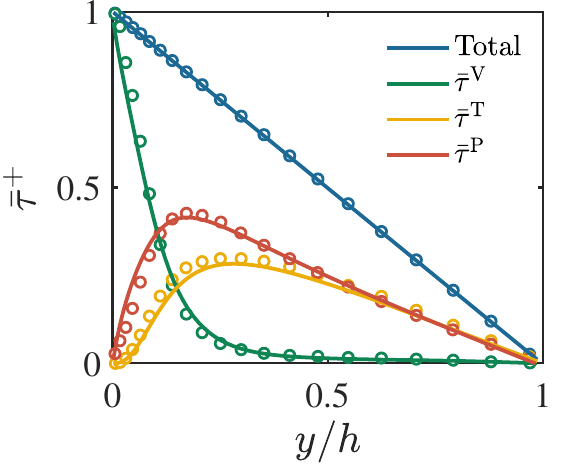}\put(-3,77){$(c)$}\end{overpic}
    \caption{\label{fig:validation-Dave}Wall-normal profiles of the $(a)$ mean flow velocity and the shear stress balance for $(b)$ $St^+=6$ and $(c)$ $St^+=30$ cases. Lines: present results; symbols: reference data adopted from \citet{Dave2023}.}
\end{figure}

Test simulations of incompressible particle-laden channel flows are performed to further demonstrate the reliability of the four-way coupled PP-DNS solver utilized in the present study.
Specifically, we replicate two cases presented in \citet{Dave2023}: channel flows of $Re_\tau=180$ laden with particles of ${\Phi}_m=1$ and $St^+=6,\,30$. 
The configurations and parameters match those in \citet{Dave2023}, where the simulations were conducted by a widely applied particle-laden flow solver, NGA \citep{Desjardins2008, Capecelatro2013}.
The bulk Mach number is set to $0.2$ to approximate the incompressible condition.
As shown in figure~\ref{fig:validation-Dave}, the mean flow velocity profiles and shear stress balance profiles obtained by the two solvers agree well with each other.
These results, along with the balanced budget profiles presented in \S~\ref{sec:results}, consolidate the reliability of the present simulations and analysis.

\bibliographystyle{jfm}
\bibliography{jfm}

\end{document}